\documentclass[aps,prl,reprint,groupedaddress,floatfix,superscriptaddress,amsmath,showpacs,longbibliography]{revtex4-1}

\usepackage{graphicx}
\usepackage{dcolumn}
\usepackage{bm}
\usepackage[colorlinks=true,linkcolor = blue,citecolor=blue,urlcolor=blue,anchorcolor = blue]{hyperref}
\usepackage{wasysym}
\usepackage{stmaryrd}
\usepackage{verbatim}
\usepackage{subfigure}
\usepackage{amsmath}
\usepackage{times}
\usepackage[version=4]{mhchem}
\usepackage{braket}
\usepackage{siunitx}
\usepackage{siunitx}
\usepackage{amssymb}
\usepackage{multirow}
\usepackage{booktabs}
\usepackage{cellspace}
\newcommand{\RNum}[1]{\uppercase\expandafter{\romannumeral #1\relax}}

\begin{document}
	
\title{Finite Temperature Magnetism in the Triangular Lattice Antiferromagnet \ce{KErTe2}}

\author{Weiwei\,Liu}
\thanks{These authors contributed to the work equally.}
\affiliation{Department of Physics, Renmin University of China, Beijing 100872, China}
\affiliation{Beijing National Laboratory for Condensed Matter Physics, Institute of Physics, Chinese Academy of Sciences, Beijing 100190, China}

\author{Zheng\,Zhang{\color{blue}$^{*}$}}
\email[e-mail:]{zhangzheng@iphy.ac.cn}
\affiliation{Beijing National Laboratory for Condensed Matter Physics, Institute of Physics, Chinese Academy of Sciences, Beijing 100190, China}

\author{Dayu\,Yan}
\affiliation{Beijing National Laboratory for Condensed Matter Physics, Institute of Physics, Chinese Academy of Sciences, Beijing 100190, China}

\author{Jianshu\,Li}
\affiliation{Beijing National Laboratory for Condensed Matter Physics, Institute of Physics, Chinese Academy of Sciences, Beijing 100190, China}

\author{Zhitao\,Zhang}
\affiliation{Anhui Province Key Laboratory of Condensed Matter Physics at Extreme Conditions, High Magnetic Field Laboratory, Chinese Academy of Sciences, Hefei 230031, China}

\author{Jianting\,Ji}
\affiliation{Beijing National Laboratory for Condensed Matter Physics, Institute of Physics, Chinese Academy of Sciences, Beijing 100190, China}

\author{Feng\,Jin}
\affiliation{Beijing National Laboratory for Condensed Matter Physics, Institute of Physics, Chinese Academy of Sciences, Beijing 100190, China}

\author{Youguo\,Shi}
\affiliation{Beijing National Laboratory for Condensed Matter Physics, Institute of Physics, Chinese Academy of Sciences, Beijing 100190, China}

\author{Qingming\,Zhang}
\affiliation{Beijing National Laboratory for Condensed Matter Physics, Institute of Physics, Chinese Academy of Sciences, Beijing 100190, China}
\affiliation{School of Physical Science and Technology, Lanzhou University, Lanzhou 730000, China}

\begin{abstract}
After the discovery of the \ce{ARECh2} (A=alkali or monovalent ions, RE=rare-earth, Ch= chalcogen) triangular lattice quantum spin liquid (QSL) family, a series of its oxide, sulfide, and selenide counterparts has been consistently reported and extensively investigated. 
While \ce{KErTe2} represents the initial synthesized telluride member, preserving its triangular spin lattice, it was anticipated that the substantial tellurium ions could impart more pronounced magnetic attributes and electronic structures to this material class.
This study delves into the magnetism of \ce{KErTe2} at finite temperatures through magnetization and electron spin resonance (ESR) measurements. 
Based on the angular momentum $\hat{J}$ after spin-orbit coupling (SOC) and symmetry analysis, we obtain the magnetic effective Hamiltonian to describe the magnetism of \ce{Er^{3+}} in R-3m space group. Applying the mean-field approximation to the Hamiltonian, we can simulate the magnetization and magnetic heat capacity of \ce{KErTe2} in paramagnetic state and determine the crystalline electric field (CEF) parameters and partial exchange interactions.
The relatively narrow energy gaps between CEF ground state and excited states exert a significant influence on the magnetism. 
For example, small CEF excitations can result in a significant broadening of the ESR linewidth at 2 K.
For the fitted exchange interactions, although the values are small, given a large angular momentum $J$ = 15/2 after SOC, they still have a noticeable effect at finite temperatures.
Notably, the heat capacity data under different magnetic fields along the c-axis direction also roughly match our calculated results, further validating the reliability of our analytical approach.
These derived parameters serve as crucial tools for future investigations into the ground state magnetism of \ce{KErTe2}.
The findings presented herein lay a foundation for the exploration of the intricate magnetism within the triangular-lattice delafossite family.

\end{abstract}

\maketitle


\emph{Introduction}---Rare-earth chalcogenides, denoted as \ce{ARECh2} (with A representing alkali or monovalent metals, RE for rare-earth, and Ch encompassing O, S, Se, and Te), have emerged as compelling candidates for quantum spin liquids (QSL)\cite{Anderson1973,ANDERSON1987,liu2018rare,bordelon2019field,ding2019gapless,ranjith2019field,baenitz2018naybs,sarkar2019quantum,Zhang2020,PhysRevB.100.224417}. 
These intriguing materials contrast with previous QSL contenders, such as \ce{ZnCu3(OH)6Cl2}\cite{PhysRevLett.98.107204}, \ce{EtMe3Sb[Pd(dmit)2]2}\cite{itou2008quantum}, and \ce{YbMgGaO4}\cite{li2015gapless,li2015rare,PhysRevLett.117.097201,PhysRevLett.117.267202,Shen2016,Paddison2016}, in that they not only possess perfect triangular structures but also offer opportunities for doping and element substitution. 
This dynamic framework proves ideal for probing intricate interactions in strongly correlated electronic systems and complex magnetism\cite{PhysRevMaterials.3.114413,PhysRevB.101.144432}, facilitated by the distinctive electronic properties intrinsic to rare-earth ions.
The potency of these materials can be attributed to several factors. 
Firstly, the robust spin-orbit coupling (SOC) in rare-earth ions engenders formidable magnetic anisotropy. 
Secondly, certain rare-earth ions, like \ce{Er^{3+}} and \ce{Yb^{3+}}, possess Kramer's degeneracy, safeguarded by time-reversal symmetry in their crystalline electric field (CEF) ground state, which is doubly degenerate. 
Thirdly, the 4f electrons of rare-earth ions, in contrast to the 3d electrons of transition metal ions, are effectively shielded by outer electrons, leading to diminished CEF excitations bridging the ground and excited states. 
Notably, these subtle energy gaps profoundly impact the thermodynamic and spectroscopic attributes of the materials, as confirmed through numerous experimental investigations.
For instance, the thermodynamic data of \ce{NaYbSe2} reveals a distinct characteristic temperature of 25 K\cite{PhysRevB.103.184419}, a manifestation of CEF effects. 
Above this temperature, CEF excitations exert significant influence on both thermodynamic and spectroscopic observations. 
Furthermore, in the case of \ce{KErSe2}, which possesses a small CEF energy gap between the ground and first excited states (approximately 1 meV)\cite{PhysRevB.101.144432}, the magnetization and thermodynamic data are predominantly governed by CEF excitations, even at low temperatures.

In a recent achievement, we have successfully synthesized single-crystal samples of \ce{KErTe2}\cite{liu2021}. 
This marks an advancement as the premier telluride among rare-earth chalcogenides to undergo comprehensive investigation. 
\ce{KErTe2} is isomorphic to \ce{NaYbCh2} (Ch encompassing O, S, Se, and Te), boasting a flawless triangular lattice structure within the R-3m space group. 
The central \ce{Er^{3+}} ion resides in an octahedral coordination, surrounded by six \ce{Te^{2-}} ligands.
Comparatively, \ce{KErTe2} deviates slightly from its counterpart \ce{KErSe2}\cite{PhysRevB.101.144432}, with only one element substituted, suggesting similar CEF excitation characteristics for these two compounds. 
It's noteworthy that the electronegativity of \ce{Te^{2-}} is comparatively less than that of \ce{Se^{2-}}, implying that the CEF excitation energy levels in \ce{KErTe2} could be even weaker than those in \ce{KErSe2}. 
Furthermore, it's important to recognize that \ce{KErTe2} displays the narrowest valence band bandwidth (approximately 0.9 eV) within this material family\cite{liu2021}, rendering it an opportune platform for delving into metallization and superconductivity phenomena\cite{zhang2020pressure,jia2020mott}.
Hence, the exploration of \ce{KErTe2}'s physical properties assumes a pivotal relevance, augmenting our understanding of this intriguing material and its potential implications.

This paper presents a comprehensive investigation of \ce{KErTe2} magnetism at finite temperatures, 
including temperature-dependent magnetization (M/H-T), magnetic field-dependent magnetization (M-H), magnetic heat capacity, and electron spin resonance (ESR).
Firstly, we have constructed an effective magnetic Hamiltonian based on the angular momentum operator $\hat{J}$ to describe the magnetism of \ce{Er^{3+}} ions, considering their weak CEF excitations.
Furthermore, in order to quantitatively analyze the measured magnetization and magnetic heat capacity data, we applied a mean-field approximation to the effective magnetic Hamiltonian. This provided a satisfactory explanation for the magnetism of \ce{KErTe2} in the paramagnetic state at high temperatures.
Next, we successfully derived the CEF parameters and angular momentum $\hat{J}$ exchange interactions $\vartheta_{zz}$ ($\sim$ 0.034 K) and $\vartheta_{\pm}$ ($\sim$ 0.001 K) for \ce{KErTe2} based on magnetization data. 
Using these parameters, we were able to effectively simulate the M/H-T and magnetic heat capacity data at finite temperatures, and compute the CEF energy levels and wavefunctions of \ce{KErTe2}.
Intriguingly, our findings reveal that \ce{KErTe2} exhibits minuscule CEF excitation energy levels, indicating that the impact of CEF excitations on magnetism cannot be ignored, even at temperatures as low as 2 K.
For the angular momentum operator $\hat{J}$ exchange interactions, $\vartheta_{zz}$ and $\vartheta_{\pm}$, although numerically small, considering the large value of $J$ = 15/2 after SOC, they can still have an impact on magnetism at finite temperatures.
Lastly, drawing on the CEF energy levels of \ce{KErTe2}, we elucidated the origin of the broad linewidth observed in the ESR spectrum at 2 K.
Our comprehensive study of \ce{KErTe2}'s CEF properties lays a solid foundation for future investigations into the material's ground state magnetism at low-temperatures. 
We propose forthcoming explorations could encompass advanced methodologies, such as thermodynamic measurements at millikelvin levels, inelastic neutron scattering experiments, exact diagonalization (ED), and density matrix renormalization group (DMRG) calculations, to deepen our comprehension of this intriguing material's properties.

\emph{Samples and Experimental Techniques}---Single crystals of \ce{KErTe2}, measuring approximately 3 mm, were successfully synthesized utilizing the Te-flux method\cite{liu2021}. The conclusive evidence of a singular high-quality crystal is substantiated through single-crystal imaging and the XRD pattern of \ce{KErTe2}\cite{liu2021, SI} (refer to Supplementary Material for details). These single crystals constituted the basis for our M/H-T, M-H, heat capacity, and ESR experiments. Furthermore, nonmagnetic isostructural \ce{KLuTe2} samples (both polycrystalline and single crystal)\cite{liu2021}, synthesized using similar methods, were used to simulate the lattice contribution to the heat capacity of \ce{KErTe2}.

Approximately 5 mg of \ce{KErTe2} single crystal was meticulously prepared to facilitate M/H-T and M-H experiments. These anisotropic measurements were conducted using a DynaCool Quantum Design physical property measurement system (PPMS) across a temperature range of 2 to 300 K, while applying magnetic fields varying from 0 to 10 T. This comprehensive study encompassed measurements both within the ab-plane and along the c-axis of the singular crystal specimen.

A sample of approximately 5 mg composed of a singular \ce{KErTe2} crystal was employed for conducting heat capacity measurements spanning the temperature range of 1.8 - 20 K. The measurements were conducted under magnetic fields of 0, 1, 2, 4, 6, and 9 T with the PPMS system. Notably, the measurements were focused along the c-axis of a single crystal.
To simulate the lattice contribution of \ce{KErTe2}, heat capacity data for a non-magnetic single crystal of \ce{KLuTe2} were also acquired using the PPMS system.

ESR measurements were meticulously conducted utilizing a Bruker EMX plus 10/12 continuous-wave spectrometer, operating within X-band frequencies (approximately $f$ $\sim$ 9.4 GHz). These measurements were carried out both within the ab-plane and along the c-axis of the \ce{KErTe2} crystal, all at a temperature of 2 K.

\begin{figure}[t]
	\includegraphics[scale=0.8]{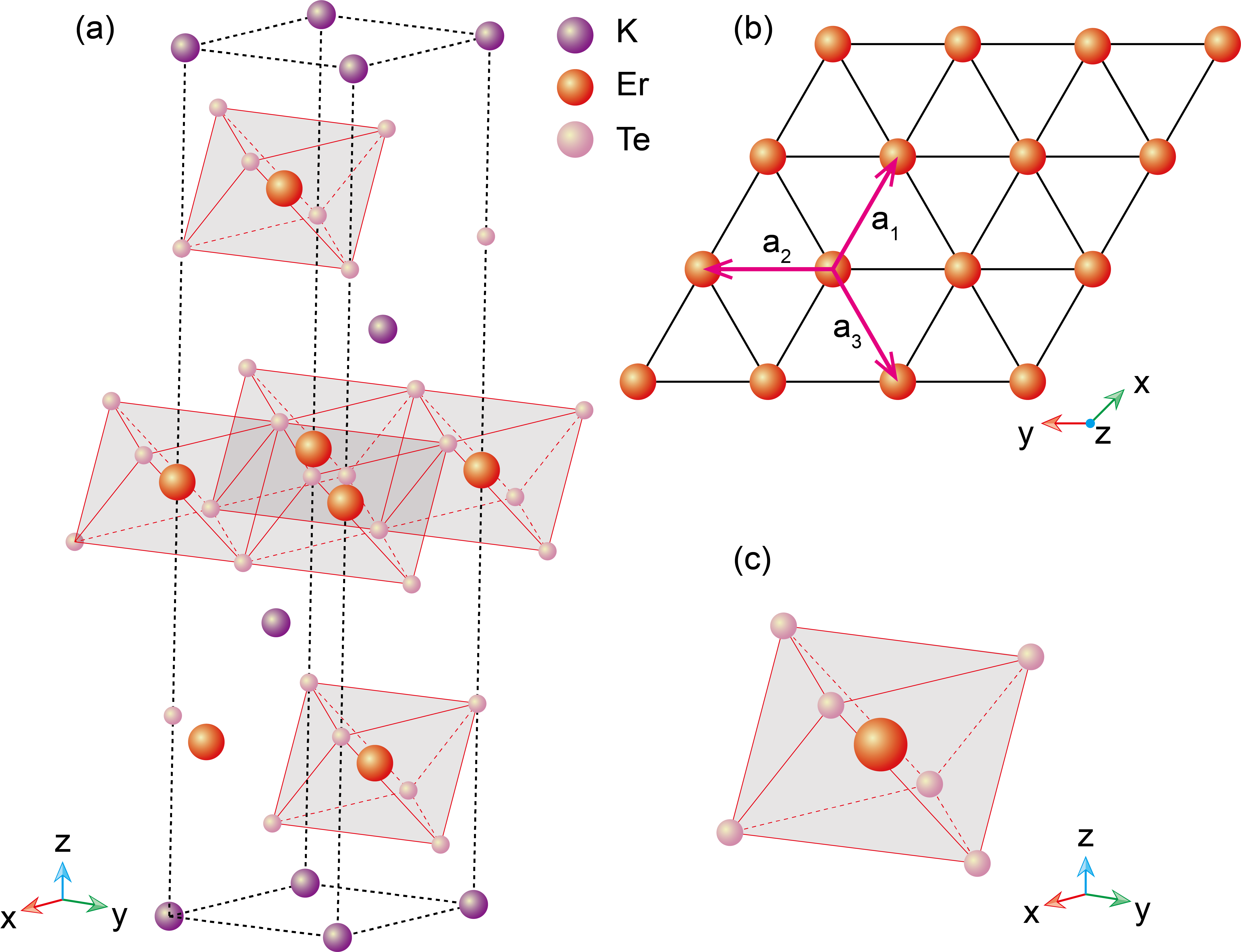}
	\caption{\label{fig:epsart}{(a) Crystal structure of \ce{KErTe2}. (b) Two-dimensional triangular plane formed by \ce{Er^{3+}} ions. Three directions, $a_{1}$, $a_{2}$, and $a_{3}$, represent the orientation of the phase factor on the anisotropic spin-exchange interaction Hamiltonian, respectively. (c) Octahedral structure with $D_{3d}$ point group symmetry formed by the central \ce{Er^{3+}} ion and the surrounding 6 \ce{Te^{2-}} anions. }}
\end{figure}


\emph{Electronic Configuration, \ce{Er^{3+}} Ion Environment, and Effective Magnetic Hamiltonian}---The $4f$ orbital of \ce{Er^{3+}} accommodates 11 electrons, yielding a total spin quantum number $S$ of 3/2. Given that the orbital quantum number $L$ equals 6 and considering the robust SOC exhibited by $4f$ electrons\cite{PhysRevB.94.035107}, there emerges a spectral term $^{4}I_{15/2}$ featuring a 16-fold degeneracy and another spectral term $^{4}I_{13/2}$ with a 14-fold degeneracy. An energy distinction of around 0.8 eV ($\sim$ 9000 K) separates these two spectral terms. Consequently, the spectral term $^{4}I_{15/2}$ denotes the SOC ground state configuration.
Furthermore, it's worth acknowledging that the energy scale of CEF excitations within rare-earth ions falls considerably below the 0.8 eV threshold. Consequently, the effect of energy level transitions stemming from SOC can be safely disregarded when assessing CEF excitation and magnetism dynamics.

Considering that the CEF excitations are typically weak in the \ce{Er^{3+}} ions system, even at several Kelvin, CEF excitations may still contribute to magnetism at low temperatures. Meanwhile, interactions between magnetic ions also gradually play a role in magnetism. Therefore, it is more appropriate to describe the magnetic Hamiltonian between \ce{Er^{3+}} ions using the angular momentum $\hat{J}$ after SOC. Based on these considerations, we have redefined the effective magnetic Hamiltonian for the \ce{Er^{3+}} ion system, incorporating CEF, angular momentum $\hat{J}$ exchange interactions, and Zeeman effects. The new Hamiltonian is described as follows\cite{PhysRevB.103.214445}:
\begin{widetext}
	\begin{equation}
		\begin{array}{l}
			H = {H_{CEF}} + \sum\limits_{\left\langle {ij} \right\rangle } {\vartheta _{ij}^{\alpha ,\beta }J_i^\alpha J_j^\beta } - {\mu _0}{\mu _B}{g_J}\sum\limits_i {{h_x}J_i^x + {h_y}J_i^y + {h_z}J_i^z} 
		\end{array}
	\end{equation}
\end{widetext}
where the first term represents the CEF Hamiltonian describing the single-site effects of the angular momentum $\hat{J}$, the second term accounts for the exchange interactions Hamiltonian capturing the many body effects of the angular momentum $\hat{J}$, $\vartheta _{ij}^{\alpha ,\beta }$ ($\alpha$, $\beta$ = $x$, $y$, or $z$) are the exchange interactions, the final term describes the Zeeman effect on the angular momentum $\hat{J}$ influenced by the magnetic field, $g_{J}$ is Landé g-factor after SOC, which is 6/5 for \ce{Er^{3+}}, and $h_{\alpha}$ ($\alpha = x, y, z$) is the magnetic field.

In \ce{KErTe2}, possessing the R-3m space group symmetry (as depicted in Fig. 1(a)), the arrangement entails an octahedral structure wherein the \ce{Er^{3+}} ion is enveloped by \ce{Te^{2-}} anions. This structure conforms to a $D_{3d}$ point group configuration (illustrated in Fig. 1(c)). Therefore, the CEF Hamiltonian can be formally expressed as follows:\cite{li2017crystalline,Zhang2020}:
\begin{widetext}
	\begin{equation}
		\begin{array}{l}
			H_{CEF} = \sum\limits_i {B_2^0O_2^0 + B_4^0O_4^0 + B_4^3O_4^3 + B_6^0O_6^0 + B_6^3O_6^3 + B_6^6O_6^6}
		\end{array}
	\end{equation}
\end{widetext}
where $B_{m}^{n}$ is the CEF parameter and $O_{m}^{n}$ is the Steven operator. Given that \ce{Er^{3+}} represents a Kramers ion characterized by an odd number of electrons, each state within the CEF is doubly degenerate and is protected by the conservation of time-reversal symmetry.
For the triangular lattice, the exchange interactions Hamiltonian for the angular momentum $\hat{J}$ can be further simplified based on symmetry analysis. The interactions along the $a_{1}$-bond allow four spatial symmetry operations: one is the $C_{2}$ rotational operation along the $a_{1}$-bond, another is the $C_{3}$ rotational operation along the c-axis, and the translations $T_{1}$ and $T_{2}$ along the $a_{1}$-bond and $a_{2}$-bond directions, respectively. Considering the $C_{2}$ rotational operation along the $a_{1}$-bond, $y$ becomes $-y$, and $z$ becomes $-z$. To keep the matrix of interactions invariant under this operation, $\vartheta_{xy}$, $\vartheta_{xz}$, $\vartheta_{yx}$, and $\vartheta_{zx}$ must be zero. As a result, the exchange interactions Hamiltonian for the $a_{1}$-bond transforms into the following form:
\begin{widetext}
	\begin{equation}
		{H_{{a_1} - bond}} = \sum\limits_{\left\langle {ij} \right\rangle } {\left( {\begin{array}{*{20}{c}}
					{J_i^x}&{J_i^y}&{J_i^z}
			\end{array}} \right)\left( {\begin{array}{*{20}{c}}
					{{\vartheta _{xx}}}&0&0\\
					0&{{\vartheta _{yy}}}&{{\vartheta _{yz}}}\\
					0&{{\vartheta _{zy}}}&{{\vartheta _{zz}}}
			\end{array}} \right)\left( {\begin{array}{*{20}{c}}
					{J_j^x}\\
					{J_j^y}\\
					{J_j^z}
			\end{array}} \right)}
	\end{equation}
\end{widetext}

Considering $T_{1}$ operation along the $a_{1}$-bond direction, position $i$ and $j$ can be interchanged. Hence, in the $a_{1}$-bond interactions matrix, $\vartheta_{yz}$ = $\vartheta_{zy}$. Angular momentum $\hat{J}$ vectors are usually represented in terms of the Hermitian Cartesian component operators, $\hat{J}^{T} = \left( J^{x} J^{y} J^{z} \right)$. Occasionally, the non-Hermitian ladder operators $J^{+} = J^{x} + iJ^{y}$ and $J^{-} = J^{x} - iJ^{y}$ are used. Under the representation of the non-Hermitian ladder operators $J^{+}$, $J^{-}$, and $J^{z}$, we define the interactions between the $J_{i}^{+}J_{j}^{+}$ and $J_{i}^{-}J_{j}^{-}$ as $\vartheta_{\pm}$, between $J_{i}^{+}J_{j}^{-}$ and $J_{i}^{-}J_{j}^{+}$ as $\vartheta_{\pm\pm}$, between $J_{i}^{+}J_{j}^{z}$, $J_{i}^{-}J_{j}^{z}$, $J_{i}^{z}J_{j}^{+}$, and $J_{i}^{z}J_{j}^{-}$ as $\vartheta_{z\pm}$, and $J_{i}^{z}J_{j}^{z}$ as $\vartheta_{zz}$. Thus,  interactions along the $a_{1}$-bond in Hamiltonian can be expressed as in the following form:
	\begin{widetext}
		\begin{equation}
			{H_{{a_1} - bond}} = \sum\limits_{\left\langle {ij} \right\rangle } {\left( {\begin{array}{*{20}{c}}
						{J_i^x}&{J_i^y}&{J_i^z}
				\end{array}} \right)\left( {\begin{array}{*{20}{c}}
						{2\left( {{\vartheta _ \pm } + {\vartheta _{ \pm  \pm }}} \right)}&0&0\\
						0&{2\left( {{\vartheta _ \pm } - {\vartheta _{ \pm  \pm }}} \right)}&{{\vartheta _{z \pm }}}\\
						0&{{\vartheta _{z \pm }}}&{{\vartheta _{zz}}}
				\end{array}} \right)\left( {\begin{array}{*{20}{c}}
						{J_j^x}\\
						{J_j^y}\\
						{J_j^z}
				\end{array}} \right)}
		\end{equation}
	\end{widetext}
Applying a clockwise 120° $C_{3}$ operation along the c-axis yields the angular momentum $\hat{J}$ interactions along the $a_{2}$-bond direction:
	\begin{widetext}
		\begin{equation}
			{H_{{a_2} - bond}} = \sum\limits_{\left\langle {ij} \right\rangle } {\left( {\begin{array}{*{20}{c}}
						{J_i^x}&{J_i^y}&{J_i^z}
				\end{array}} \right)\left( {\begin{array}{*{20}{c}}
						{2{\vartheta _ \pm } - {\vartheta _{ \pm  \pm }}}&{\sqrt 3 {\vartheta _{ \pm  \pm }}}&{\frac{{\sqrt 3 }}{2}{\vartheta _{z \pm }}}\\
						{\sqrt 3 {\vartheta _{ \pm  \pm }}}&{2{\vartheta _ \pm } + {\vartheta _{ \pm  \pm }}}&{ - \frac{1}{2}{\vartheta _{z \pm }}}\\
						{\frac{{\sqrt 3 }}{2}{\vartheta _{z \pm }}}&{ - \frac{1}{2}{\vartheta _{z \pm }}}&{{\vartheta _{zz}}}
				\end{array}} \right)\left( {\begin{array}{*{20}{c}}
						{J_j^x}\\
						{J_j^y}\\
						{J_j^z}
				\end{array}} \right)}
		\end{equation}
	\end{widetext}
Continuing the clockwise rotation by 120° further yields the angular momentum $\hat{J}$ interactions Hamiltonian along the $a_{3}$-bond direction:
	\begin{widetext}
		\begin{equation}
			{H_{{a_3} - bond}} = \sum\limits_{\left\langle {ij} \right\rangle } {\left( {\begin{array}{*{20}{c}}
						{J_i^x}&{J_i^y}&{J_i^z}
				\end{array}} \right)\left( {\begin{array}{*{20}{c}}
						{2{\vartheta _ \pm } - {\vartheta _{ \pm  \pm }}}&{ - \sqrt 3 {\vartheta _{ \pm  \pm }}}&{ - \frac{{\sqrt 3 }}{2}{\vartheta _{z \pm }}}\\
						{ - \sqrt 3 {\vartheta _{ \pm  \pm }}}&{2{\vartheta _ \pm } + {\vartheta _{ \pm  \pm }}}&{ - \frac{1}{2}{\vartheta _{z \pm }}}\\
						{ - \frac{{\sqrt 3 }}{2}{\vartheta _{z \pm }}}&{ - \frac{1}{2}{\vartheta _{z \pm }}}&{{\vartheta _{zz}}}
				\end{array}} \right)\left( {\begin{array}{*{20}{c}}
						{J_j^x}\\
						{J_j^y}\\
						{J_j^z}
				\end{array}} \right)}
		\end{equation}
	\end{widetext}

Based on the effective magnetic Hamiltonian established for \ce{Er^{3+}} ions, we can analyze the magnetization and magnetic heat capacity data of \ce{KErTe2} at finite temperatures.

If we consider the influence of the magnetic field on the \ce{Er^{3+}} ions, Zeeman term $H_{zeeman}$ should be incorporated into the effective magnetic Hamiltonian.
\begin{equation}
	H_{zeeman} = -\mu_{0}\mu_B g_{j} \sum_{i} h_x \hat{J}_{i}^{x} + h_y \hat{J}_{i}^{y} + h_z \hat{J}_{i}^{z}
\end{equation}
where $g_{j}$ is Lande g-factor after the SOC, which is 6/5 for \ce{Er^{3+}}, and $h_{\alpha}$ ($\alpha = x, y, z$) is the magnetic field along the different directions.
Since the magnetic effective Hamiltonian is constructed based on the total angular momentum $J = 15/2$, there is no need to consider the anisotropy of the g-factor.

\begin{figure*}[t]
	\includegraphics[scale=0.85]{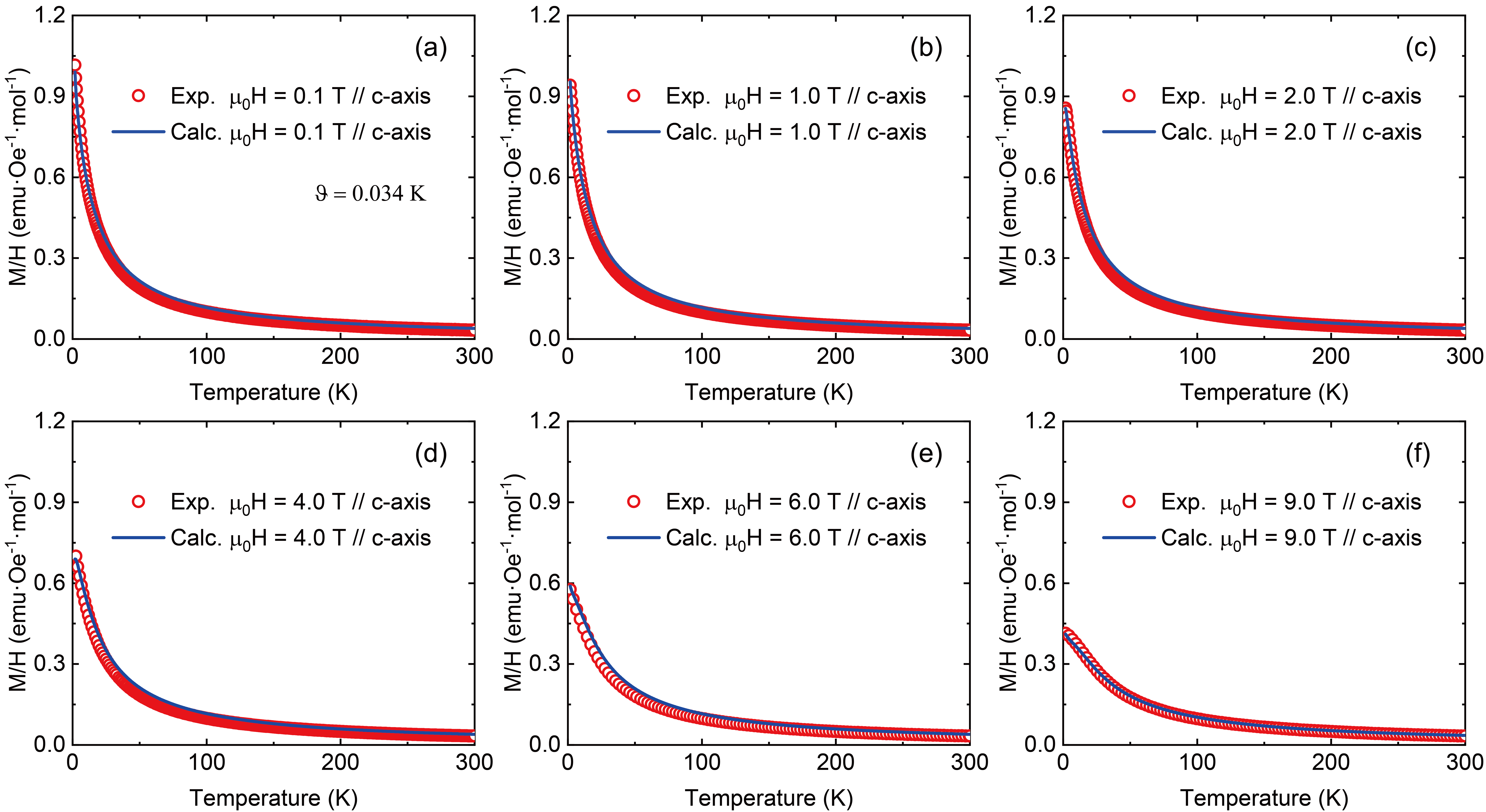}
	\caption{\label{fig:epsart}{M/H-T data (open red circles) under different magnetic fields along the c-axis of \ce{KErTe2} and the simulation results (solid blue lines).}}
\end{figure*}
\emph{M/H-T, M-H, and magnetic heat capacity of \ce{KErTe2}}---A thorough comprehension of the CEF excitations and exchange interactions in rare-earth magnetic materials serves as a cornerstone for advancing investigations into low-energy quantum magnetism, particularly within the realm of rare-earth QSL candidate materials. 
Among the various experimental techniques, inelastic neutron scattering (INS) is widely regarded as the most precise method for discerning CEF excitations and the associated CEF parameters in rare-earth materials. 
Besides, the low-energy excitations in INS also contain information about exchange interactions. By analyzing these excitation spectra, exchange interactions can be deduced.
In addition to INS, magnetization and magnetic heat capacity measurements also yield valuable insights into the CEF and exchange interactions. 
Specifically, in the case of \ce{KErTe2}, 
the magnetization and heat capacity data under different magnetic fields reveal the response of the system to CEF effects and exchange interactions.
By employing the angular momentum $\hat{J}$ magnetic effective Hamiltonian and relevant magnetization and thermodynamic calculation formulas, we can simulate the magnetization and magnetic heat capacity data, and obtain these parameters. 
The material under investigation in this study, \ce{KErTe2}, is characterized by several key attributes. 
Firstly, \ce{KErTe2} shares significant similarities with \ce{KErSe2}.
In the case of \ce{KErSe2}, INS data has revealed an energy gap of approximately 1 meV from the CEF ground state to the first excited state\cite{PhysRevB.101.144432}. 
Considering the weaker electronegativity of \ce{Te^{2-}} compared to \ce{Se^{2-}}, it is conceivable that \ce{KErTe2} may possess even smaller CEF excitations.
A notable example involves \ce{NaYbO2}, \ce{NaYbS2}, and \ce{NaYbSe2}. INS indicates that the CEF first excitation energy levels of these materials are 34.8 meV\cite{ding2019gapless}, 23 meV\cite{baenitz2018naybs}, and 15.79 meV\cite{Zhang2020}, respectively. It is clearly evident that as the electronegativity of O, S, and Se decrease, the CEF first excitation energy level also continuously decreases.
Secondly, the temperature range of M/H-T measurements is 1.8 $\sim$ 300 K, and for heat capacity measurements, it is 1.8 $\sim$ 20 K. This range effectively encompasses the magnetic and thermodynamic information related to the CEF excitations of \ce{KErTe2}.
In relation to this, we have measured the heat capacity data of \ce{KLuTe2}, applying Debye's formula to accurately simulate the lattice heat capacity of \ce{KErTe2}. 
Thirdly, the smooth M/H-T data indicate that there are no magnetic phase transitions occurring in \ce{KErTe2} within the temperature range of 1.8 to 300 K.
Therefore, by employing mean-field approximation on the effective magnetic Hamiltonian, the magnetism in this paramagnetic state can be well described.

We briefly describe the mean-field Hamiltonian suitable for the paramagnetic state. When the magnetic field is applied along the c-axis, the magnetic moment of angular momentum $\hat{J}$ is almost aligned with the direction of the c-axis. The contributions of order parameters $\left\langle J^{x} \right\rangle$ and $\left\langle J^{y} \right\rangle$ can be neglected. Thus, the Hamiltonian can be simplified to the following form:
	\begin{widetext}
		\begin{equation}
			{H_{MF - c - axis}} = \sum\limits_i {B_2^0O_2^0 + B_4^0O_4^0 + B_4^3O_4^3 + B_6^0O_6^0 + B_6^3O_6^3 + B_6^6O_6^6}  + \frac{{6{\vartheta _ \pm }{M^z}}}{{{\mu _0}{\mu _B}{g_J}}}\sum\limits_i {J_i^z}  - {\mu _0}{\mu _B}{g_J}\sum\limits_i {{h_z}J_i^z} 
		\end{equation}
	\end{widetext}
Similarly, when the magnetic field is in the ab-plane, the contribution of the order parameter $\left\langle J^{z} \right\rangle$ is almost zero. The Hamiltonian takes the following form:
\begin{widetext}
	\begin{equation}
		{H_{MF - ab - plane}} = \sum\limits_i {B_2^0O_2^0 + B_4^0O_4^0 + B_4^3O_4^3 + B_6^0O_6^0 + B_6^3O_6^3 + B_6^6O_6^6}  + \frac{{12{\vartheta _ \pm }{M^x}}}{{{\mu _0}{\mu _B}{g_J}}}\sum\limits_i {J_i^x}  - {\mu _0}{\mu _B}{g_J}\sum\limits_i {{h_x}J_i^x}
	\end{equation}
\end{widetext}
Therefore, under the mean-field approximation, the CEF parameters and the diagonal exchange interactions $\vartheta_{zz}$ and $\vartheta_{\pm}$ are retained. For the remaining two off-diagonal interactions, $\vartheta_{\pm\pm}$ and $\vartheta_{z\pm}$, the ED or DMRG methods are employed. Meanwhile, thermodynamic or INS data at lower temperatures are required to extract the off-diagonal interactions.

In light of these considerations, a promising avenue emerges: 
fitting and simulating the magnetization and magnetic heat capacity data of \ce{KErTe2} to obtain the CEF parameters and interactions $\vartheta_{zz}$ and $\vartheta _ \pm$.
This approach allows us to gain a comprehensive understanding of the magnetism of \ce{KErTe2} at finite temperatures.

\begin{figure*}[t]
	\includegraphics[scale=0.85]{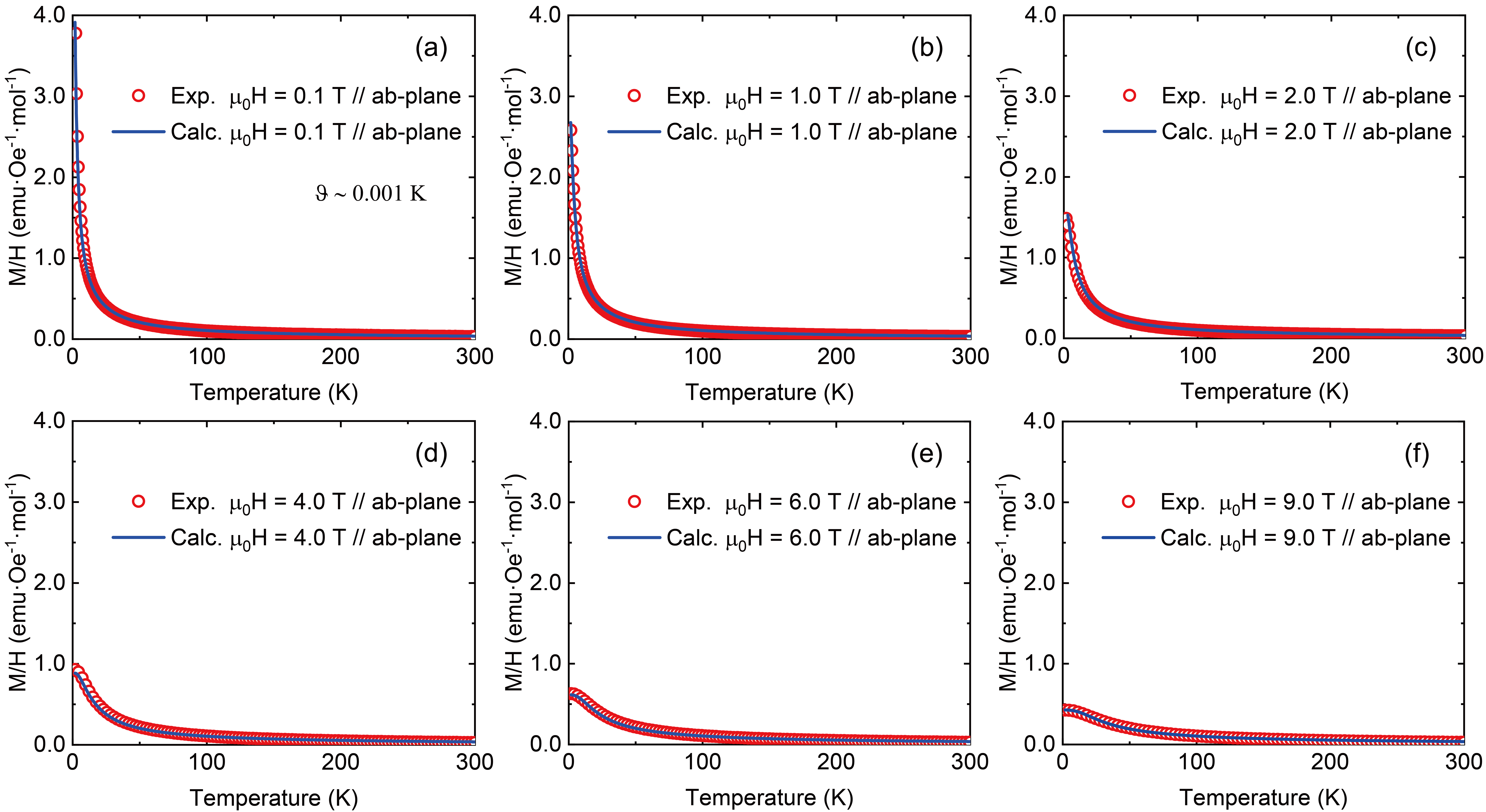}
	\caption{\label{fig:epsart}{M/H-T data (open red circles) under different magnetic fields in the ab-plane of \ce{KErTe2} and the simulation results (solid blue lines).}}
\end{figure*}

For the purpose of fitting the CEF parameters and exchange interactions $\vartheta_{zz}$ and $\vartheta _ \pm$, we have opted to utilize the M/H-T data of \ce{KErTe2} under a magnetic field 0.1 T. 
This fitting process encompasses a temperature range spanning from 1.8 to 300 K. As a starting point for our parameter estimation, we have chosen to employ the CEF parameters obtained from \ce{KErSe2} as initial values\cite{PhysRevB.101.144432}. The formula used for magnetization calculations to facilitate the CEF parameter and exchange interactions fitting is expressed as follows:
\begin{equation}
	\chi _{ab - plane} = \frac{{{\mu _B}{g_{j}}{N_A}\sum\limits_i {\exp \left( { - \frac{{{E_i}}}{{{k_B}T}}} \right)\left\langle {{\varphi _i}} \right|{J_x}\left| {{\varphi _i}} \right\rangle } }}{{{h_{ab}}\sum\limits_i {\exp \left( { - \frac{{{E_i}}}{{{k_B}T}}} \right)} }}
\end{equation}
\begin{equation}
	\chi _{c - axis} = \frac{{{\mu _B}{g_{j}}{N_A}\sum\limits_i {\exp \left( { - \frac{{{E_i}}}{{{k_B}T}}} \right)\left\langle {{\varphi _i}} \right|{J_z}\left| {{\varphi _i}} \right\rangle } }}{{{h_c}\sum\limits_i {\exp \left( { - \frac{{{E_i}}}{{{k_B}T}}} \right)} }}
\end{equation}
where $h_{ab}$ and $h_{c}$ are magnetic fields in the ab-plane and along the c-axis, respectively. $E_{i}$ and $\left| {{\varphi _i}} \right\rangle$ are the energy level and wave function obtained by the diagonalized mean-field Hamiltonian.

Fig. 2 and Fig. 3 display the M/H-T experimental data of \ce{KErTe2} under different magnetic fields, along with the corresponding simulation results. Through the fitting, we obtained the CEF parameters (see Table I) and $\vartheta_{zz}$ = 0.034 K  and $\vartheta _ \pm$ = 0.001 K. From the fitting and simulation results, it is evident that the CEF parameters and exchange interactions obtained from our fitting can effectively reproduce the M/H-T data under different magnetic fields, indicating the reasonableness of these parameters. We also calculated the CEF energy levels and wave functions, as shown in Table II.
For the exchange interactions, although the fitted parameters are tiny, considering the large angular momentum of $J=15/2$ after SOC, they would be amplified by a factor of $\frac{J(J+1)}{S(S+1)} = 85$ compared to the effective $S=1/2$\cite{PhysRevB.101.144432}.
Similarly, based on the molecular-field theory, the spin exchange interactions for \ce{KErSe2} derived form fitting magnetization data of are $J_{ab} = 0.4(3)$ $\mu$eV ($\sim$ 0.005 K) and $J_{c}$ = -2.4(4) $\mu$eV ($\sim$ -0.028 K)\cite{PhysRevB.101.144432}, which are on the same energy scale to the spin exchange interactions of \ce{KErTe2}.
\begin{table*}
	\caption{The fitted results of \ce{KErTe2} CEF parameters}
	\begin{ruledtabular}
		\begin{tabular}{cccccc}
			$B_{2}^{0}$ (meV) & $B_{4}^{0}$ (meV) & $B_{4}^{3}$ (meV) & $B_{6}^{0}$ (meV) & $B_{6}^{3}$ (meV) & $B_{6}^{6}$ (meV)\\
			\hline
			-3.070$\times$10$^{-2}$ & -1.942$\times$10$^{-2}$ & -1.050$\times$10$^{-2}$ & 1.378$\times$10$^{-6}$ & 6.146$\times$10$^{-6}$ & 5.536$\times$10$^{-5}$ \\
		\end{tabular}
	\end{ruledtabular}
\end{table*}

\begin{table*}
	\renewcommand{\arraystretch}{1.3}
	\caption{The CEF energy levels and wave functions of \ce{KErTe2}}
	\begin{ruledtabular}
		\begin{tabular}{ccccccc}
			& Energy levels (meV) & \multicolumn{5}{c}{CEF wave functions} \\
			\hline
			
			Ground State  & 0 & \multicolumn{5}{l}{$\begin{array}{l}
						\left| {{\psi ^ + }} \right\rangle  = 0.17\left| {\frac{{13}}{2}} \right\rangle  + 0.29\left| {\frac{{11}}{2}} \right\rangle  + 0.18\left| {\frac{7}{2}} \right\rangle  + 0.50\left| {\frac{5}{2}} \right\rangle  + 0.03\left| {\frac{1}{2}} \right\rangle \\
						+ 0.10\left| { - \frac{1}{2}} \right\rangle  - 0.16\left| { - \frac{5}{2}} \right\rangle  - 0.54\left| { - \frac{7}{2}} \right\rangle  + 0.09\left| { - \frac{{11}}{2}} \right\rangle  + 0.52\left| { - \frac{{13}}{2}} \right\rangle \\
						\\
						\left| {{\psi ^ - }} \right\rangle  = 0.52\left| {\frac{{13}}{2}} \right\rangle  - 0.09\left| {\frac{{11}}{2}} \right\rangle  + 0.54\left| {\frac{7}{2}} \right\rangle  - 0.16\left| {\frac{5}{2}} \right\rangle  + 0.10\left| {\frac{1}{2}} \right\rangle \\
						- 0.03\left| { - \frac{1}{2}} \right\rangle  - 0.50\left| { - \frac{5}{2}} \right\rangle  - 0.18\left| { - \frac{7}{2}} \right\rangle  + 0.29\left| { - \frac{{11}}{2}} \right\rangle  - 0.17\left| { - \frac{{13}}{2}} \right\rangle 
					\end{array}$} \\
			    
			\hline
			
			1st Excited State  & 0.49 & \multicolumn{5}{l}{$\begin{array}{l}
					\left| {{\psi ^ + }} \right\rangle  = 0.02\left| {\frac{{15}}{2}} \right\rangle  + 0.16\left| {\frac{9}{2}} \right\rangle  + 0.47\left| {\frac{3}{2}} \right\rangle  + 0.17\left| { - \frac{3}{2}} \right\rangle  - 0.54\left| { - \frac{9}{2}} \right\rangle  + 0.65\left| { - \frac{{15}}{2}} \right\rangle \\
					\\
					\left| {{\psi ^ - }} \right\rangle  =  - 0.65\left| {\frac{{15}}{2}} \right\rangle  - 0.54\left| {\frac{9}{2}} \right\rangle  - 0.17\left| {\frac{3}{2}} \right\rangle  + 0.47\left| { - \frac{3}{2}} \right\rangle  - 0.16\left| { - \frac{9}{2}} \right\rangle  + 0.02\left| { - \frac{{15}}{2}} \right\rangle 
				\end{array}$} \\
			
			\hline
			
			2nd Excited State  & 2.72 & \multicolumn{5}{l}{$\begin{array}{l}
					\left| {{\psi ^ + }} \right\rangle  = 0.09\left| {\frac{{13}}{2}} \right\rangle  - 0.40\left| {\frac{{11}}{2}} \right\rangle  - 0.09\left| {\frac{7}{2}} \right\rangle  + 0.07\left| {\frac{5}{2}} \right\rangle  - 0.29\left| {\frac{1}{2}} \right\rangle \\
					+ 0.78\left| { - \frac{1}{2}} \right\rangle  + 0.03\left| { - \frac{5}{2}} \right\rangle  - 0.23\left| { - \frac{7}{2}} \right\rangle  + 0.15\left| { - \frac{{11}}{2}} \right\rangle  - 0.24\left| { - \frac{{13}}{2}} \right\rangle \\
					\\
					\left| {{\psi ^ - }} \right\rangle  = 0.24\left| {\frac{{13}}{2}} \right\rangle  + 0.15\left| {\frac{{11}}{2}} \right\rangle  - 0.23\left| {\frac{7}{2}} \right\rangle  - 0.03\left| {\frac{5}{2}} \right\rangle  - 0.78\left| {\frac{1}{2}} \right\rangle \\
					- 0.29\left| { - \frac{1}{2}} \right\rangle  + 0.07\left| { - \frac{5}{2}} \right\rangle  + 0.09\left| { - \frac{7}{2}} \right\rangle  + 0.40\left| { - \frac{{11}}{2}} \right\rangle  + 0.09\left| { - \frac{{13}}{2}} \right\rangle 
				\end{array}$} \\
			
			\hline
			
			3rd Excited State  & 4.62 & \multicolumn{5}{l}{$\begin{array}{l}
					\left| {{\psi ^ + }} \right\rangle  = 0.45\left| {\frac{{13}}{2}} \right\rangle  - 0.40\left| {\frac{{11}}{2}} \right\rangle  + 0.08\left| {\frac{7}{2}} \right\rangle  - 0.36\left| {\frac{5}{2}} \right\rangle  - 0.05\left| {\frac{1}{2}} \right\rangle \\
					- 0.05\left| { - \frac{1}{2}} \right\rangle  + 0.33\left| { - \frac{5}{2}} \right\rangle  - 0.08\left| { - \frac{7}{2}} \right\rangle  - 0.37\left| { - \frac{{11}}{2}} \right\rangle  + 0.49\left| { - \frac{{13}}{2}} \right\rangle \\
					\\
					\left| {{\psi ^ - }} \right\rangle  = 0.48\left| {\frac{{13}}{2}} \right\rangle  + 0.37\left| {\frac{{11}}{2}} \right\rangle  + 0.08\left| {\frac{7}{2}} \right\rangle  + 0.33\left| {\frac{5}{2}} \right\rangle  - 0.05\left| {\frac{1}{2}} \right\rangle \\
					+ 0.05\left| { - \frac{1}{2}} \right\rangle  + 0.36\left| { - \frac{5}{2}} \right\rangle  + 0.08\left| { - \frac{7}{2}} \right\rangle  - 0.40\left| { - \frac{{11}}{2}} \right\rangle  - 0.45\left| { - \frac{{13}}{2}} \right\rangle 
				\end{array}$} \\
			
			\hline
			
			4th Excited State  & 4.67 & \multicolumn{5}{l}{$\begin{array}{l}
					\left| {{\psi ^ + }} \right\rangle  = 0.71\left| {\frac{{15}}{2}} \right\rangle  - 0.22\left| {\frac{9}{2}} \right\rangle  - 0.32\left| {\frac{3}{2}} \right\rangle  + 0.55\left| { - \frac{3}{2}} \right\rangle  - 0.18\left| { - \frac{9}{2}} \right\rangle  - 0.02\left| { - \frac{{15}}{2}} \right\rangle \\
					\\
					\left| {{\psi ^ - }} \right\rangle  =  - 0.02\left| {\frac{{15}}{2}} \right\rangle  + 0.18\left| {\frac{9}{2}} \right\rangle  + 0.55\left| {\frac{3}{2}} \right\rangle  + 0.32\left| { - \frac{3}{2}} \right\rangle  - 0.22\left| { - \frac{9}{2}} \right\rangle  - 0.71\left| { - \frac{{15}}{2}} \right\rangle 
				\end{array}$} \\
			
			\hline
			
			5th Excited State  & 18.43 & \multicolumn{5}{l}{$\begin{array}{l}
					\left| {{\psi ^ + }} \right\rangle  = 0.29\left| {\frac{{13}}{2}} \right\rangle  - 0.26\left| {\frac{{11}}{2}} \right\rangle  - 0.35\left| {\frac{7}{2}} \right\rangle  + 0.16\left| {\frac{5}{2}} \right\rangle  + 0.48\left| {\frac{1}{2}} \right\rangle \\
					- 0.25\left| { - \frac{1}{2}} \right\rangle  + 0.31\left| { - \frac{5}{2}} \right\rangle  - 0.18\left| { - \frac{7}{2}} \right\rangle  + 0.50\left| { - \frac{{11}}{2}} \right\rangle  - 0.15\left| { - \frac{{13}}{2}} \right\rangle \\
					\\
					\left| {{\psi ^ - }} \right\rangle  =  - 0.15\left| {\frac{{13}}{2}} \right\rangle  - 0.50\left| {\frac{{11}}{2}} \right\rangle  + 0.18\left| {\frac{7}{2}} \right\rangle  + 0.31\left| {\frac{5}{2}} \right\rangle  - 0.25\left| {\frac{1}{2}} \right\rangle \\
					- 0.48\left| { - \frac{1}{2}} \right\rangle  - 0.16\left| { - \frac{5}{2}} \right\rangle  - 0.35\left| { - \frac{7}{2}} \right\rangle  - 0.26\left| { - \frac{{11}}{2}} \right\rangle  - 0.29\left| { - \frac{{13}}{2}} \right\rangle 
				\end{array}$} \\
			
			\hline
			
			6th Excited State  & 20.23 & \multicolumn{5}{l}{$\begin{array}{l}
					\left| {{\psi ^ + }} \right\rangle  =  - 0.26\left| { - \frac{{15}}{2}} \right\rangle  + 0.77\left| { - \frac{9}{2}} \right\rangle  - 0.48\left| { - \frac{3}{2}} \right\rangle  + 0.32\left| {\frac{3}{2}} \right\rangle  - 0.10\left| {\frac{9}{2}} \right\rangle \\
					\\
					\left| {{\psi ^ - }} \right\rangle  = 0.10\left| { - \frac{9}{2}} \right\rangle  + 0.32\left| { - \frac{3}{2}} \right\rangle  + 0.48\left| {\frac{3}{2}} \right\rangle  + 0.77\left| {\frac{9}{2}} \right\rangle  + 0.26\left| {\frac{{15}}{2}} \right\rangle 
				\end{array}$} \\
			
			\hline
			
			7th Excited State  & 21.43 & \multicolumn{5}{l}{$\begin{array}{l}
					\left| {{\psi ^ + }} \right\rangle  = 0.06\left| {\frac{{13}}{2}} \right\rangle  + 0.32\left| {\frac{{11}}{2}} \right\rangle  - 0.14\left| {\frac{7}{2}} \right\rangle  - 0.58\left| {\frac{5}{2}} \right\rangle  + 0.01\left| {\frac{1}{2}} \right\rangle \\
					- 0.07\left| { - \frac{1}{2}} \right\rangle  - 0.13\left| { - \frac{5}{2}} \right\rangle  - 0.66\left| { - \frac{7}{2}} \right\rangle  - 0.07\left| { - \frac{{11}}{2}} \right\rangle  - 0.28\left| { - \frac{{13}}{2}} \right\rangle \\
					\\
					\left| {{\psi ^ - }} \right\rangle  =  - 0.28\left| {\frac{{13}}{2}} \right\rangle  + 0.07\left| {\frac{{11}}{2}} \right\rangle  + 0.66\left| {\frac{7}{2}} \right\rangle  - 0.13\left| {\frac{5}{2}} \right\rangle  - 0.07\left| {\frac{1}{2}} \right\rangle \\
					- 0.01\left| { - \frac{1}{2}} \right\rangle  + 0.58\left| { - \frac{5}{2}} \right\rangle  - 0.14\left| { - \frac{7}{2}} \right\rangle  + 0.32\left| { - \frac{{11}}{2}} \right\rangle  - 0.06\left| { - \frac{{13}}{2}} \right\rangle 
				\end{array}$} \\
		\end{tabular}
	\end{ruledtabular}
\end{table*}

\begin{table*}
	\caption{The CEF energy levels of \ce{KErTe2} and \ce{KErSe2}\cite{PhysRevB.101.144432}}
	\begin{ruledtabular}
		\begin{tabular}{ccccccccc}
			            & CEF0 (meV) & CEF1 (meV) & CEF2 (meV) & CEF3 (meV) & CEF4 (meV) & CEF5 (meV) & CEF6 (meV) & CEF7 (meV) \\
			\hline
			\ce{KErTe2} & 0 & 0.49 & 2.72 & 4.62 & 4.67 & 18.43 & 20.23 & 21.43 \\
			\hline
			\ce{KErSe2} & 0 & 0.903 & 3.491 & 5.134 & 5.538 & 23.3 & 25.2 & 25.4 
		\end{tabular}
	\end{ruledtabular}
\end{table*}

Basically, heat capacity reflects the energy spectrum of microscopic states. Therefore, magnetic heat capacity can be used to study the magnetic excitation information present at finite temperatures. The thermodynamic calculation formula to calculate magnetic heat capacity is given as follows:
\begin{equation}
	\begin{array}{l}
		{C_{mag}} = \frac{1}{{{k_B}{T^2}}}\frac{\partial }{{\partial \beta }}\left( {\frac{1}{Z}\frac{{\partial Z}}{{\partial \beta }}} \right)\\
		= \frac{1}{{{k_B}{T^2}}}\left( {\frac{1}{Z}\sum\limits_n {E_n^2{{\mathop{\rm e}\nolimits} ^{ - \beta {E_n}}} - {{\left[ {\frac{1}{Z}\sum\limits_n {{E_n}{{\mathop{\rm e}\nolimits} ^{ - \beta {E_n}}}} } \right]}^2}} } \right)
	\end{array}
\end{equation}
where $Z$ is the partition function. Similarly, we calculated the magnetic heat capacity data under different magnetic fields based on the mean-field Hamiltonian and compared them with experimental results, as shown in Fig. 4. Our simulated results are in good agreement with experimental measurements, especially in the high-temperature and high-magnetic-field ranges. Meanwhile, we can put forward the following discussion.
\begin{figure*}[t]
	\includegraphics[scale=0.85]{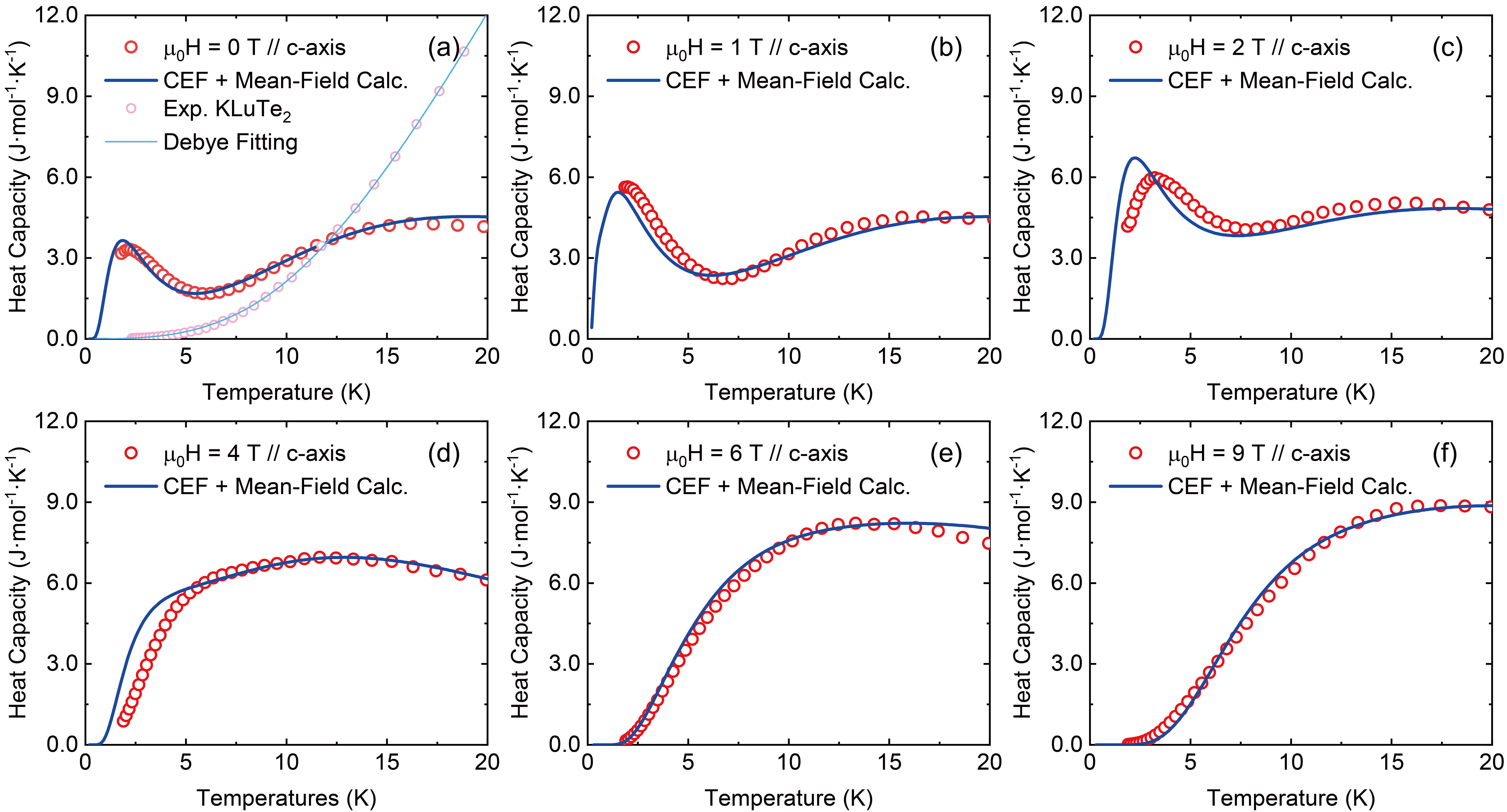}
	\caption{\label{fig:epsart}{The red open circles represent experimental results of the magnetic heat capacity under different magnetic fields. The solid blue lines are the simulation results based on the CEF and mean-field theory. Inset: The heat capacity of the non-magnetic reference sample of \ce{KLuTe2}. The light red open circles are the experimental data, while the light blue solid line is the fitting results based on the Debye formula.}}
\end{figure*}

(I) We use the heat capacity of the non-magnetic reference sample of \ce{KLuTe2} to simulate the lattice contribution to the heat capacity of \ce{KErTe2}, as shown in Fig. 4(a). When the temperature is below 20 K, the heat capacity of \ce{KLuTe2} can be well fitted by the Debye formula. By subtracting the lattice heat capacity contribution, we obtain the magnetic heat capacity data of \ce{KErTe2} under different magnetic fields along the c-axis. For the magnetic heat capacity data under the magnetic field range of 0 to 2 T, we observed a distinct broad peak within the temperature range of 1.8 to 5 K, and with the increasing magnetic field, the magnetic heat capacity contribution within this temperature range significantly increases. This is primarily attributed to the combined effect of low-energy CEF excitations and angular momentum $\hat{J}$ exchange interactions. Furthermore, the energy levels of these doubly degenerate CEF can be released by magnetic fields, allowing for the release of more magnetic entropy. Therefore, we observe a significant enhancement of this broad heat capacity peak.

(II) After the magnetic field exceeds 4 T, the magnetic heat capacity becomes smoother in the temperature range of 1.8 to 20 K. The main reason for this is that the magnetic moments of the magnetic system have already saturated after 4 T. As shown in Fig. 5, the magnetic system enters a state of magnetic moment saturation after approximately 2 T.

(III) Through comparison between simulated results and experimental data, we found that under high-temperature and high magnetic field conditions, the calculated results agree with the experimental data. However, discrepancies between the calculated results and experimental data exist in the low-temperature and weak-field range. This is because heat capacity describes the microstates, and as the temperature decrease, the off-diagonal interactions between angular momentum $\hat{J}$ and quantum fluctuations become more significant. The mean-field approximation we used only retains the diagonal interactions, hence it is insufficient to describe the microstates at low temperatures.

(IV) We also compare the CEF energy levels of \ce{KErTe2} and \ce{KErSe2}\cite{PhysRevB.101.144432}. In Table III, we can clearly see that CEF excitations in \ce{KErTe2} are overall weaker than those in \ce{KEeSe2}. This corresponds with the observation that the intensity of CEF excitations diminishes as the electronegativity of the coordinating anions decreases.

\begin{figure}[b]
	\includegraphics[scale=0.95]{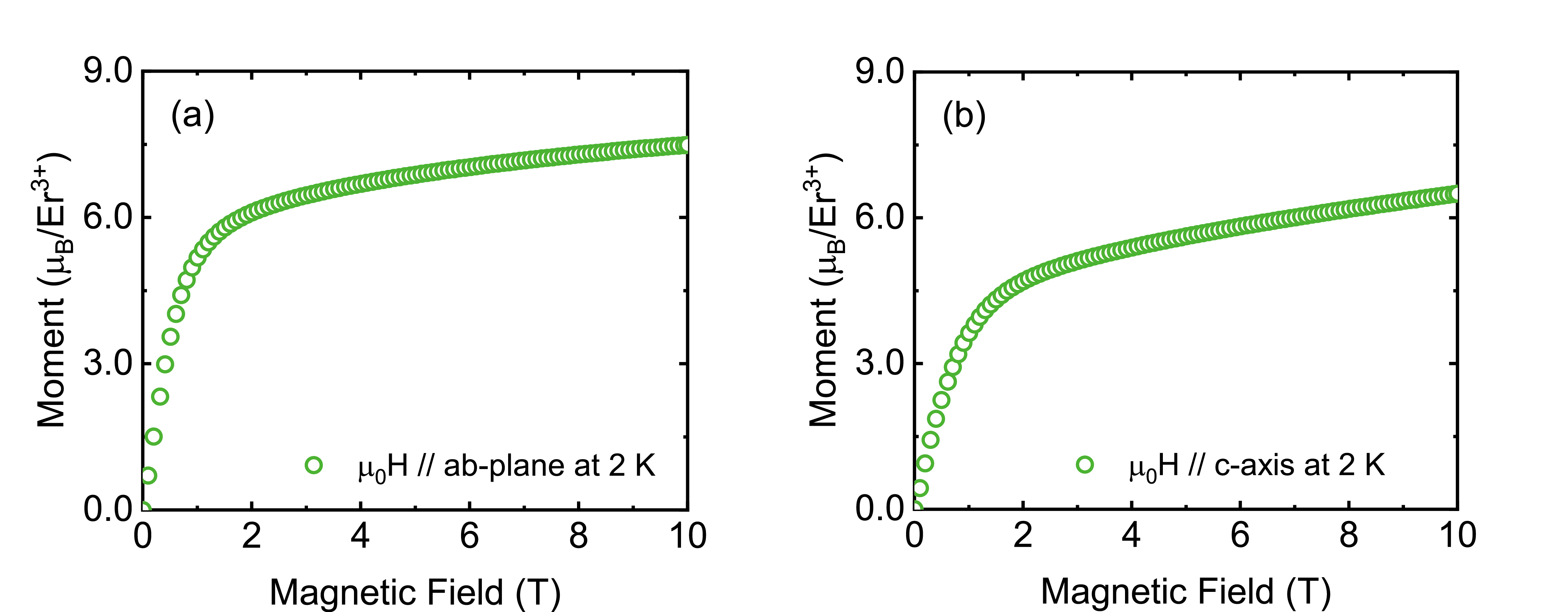}
	\caption{\label{fig:epsart}{(a) M-H data of the magnetic field in the ab-plane at 2 K. (b) M-H data of the magnetic field along the c-axis at 2 K.}}
\end{figure}

Furthermore, let's briefly discuss the magnetic anisotropy of \ce{KErTe2}.
The magnetic anisotropy is particularly pronounced at 0.1 T and gradually diminishes as the magnetic field increases. This phenomenon becomes more evident when examining the magnetization data at 2 K (see Fig. 5).
We noticed that the magnetic moment saturates after approximately 2 T. The ratio of the magnetic moment within the ab-plane to that along the c-axis stands at approximately 1.2:1.
Ignoring the Van Vleck paramagnetism contributed by the CEF, the ratio can be used to discuss the magnetic anisotropy of \ce{KErTe2}.
Compared with the ratio of 1.4:1 in \ce{KErSe2}\cite{PhysRevB.101.144432}, the magnetic anisotropy of \ce{KErTe2} is weaker. This implies that chalcogenide anions can affect the magnetic anisotropy. Similar phenomena are also observed in other compounds, such as \ce{NaYbS2}\cite{sichelschmidt2019electron} and \ce{NaYbSe2}\cite{Sichelschmidt2020,Zhang2020}, where \ce{Yb^{3+}} serves as the magnetic ion.

\begin{figure}[b]
	\includegraphics[scale=0.44]{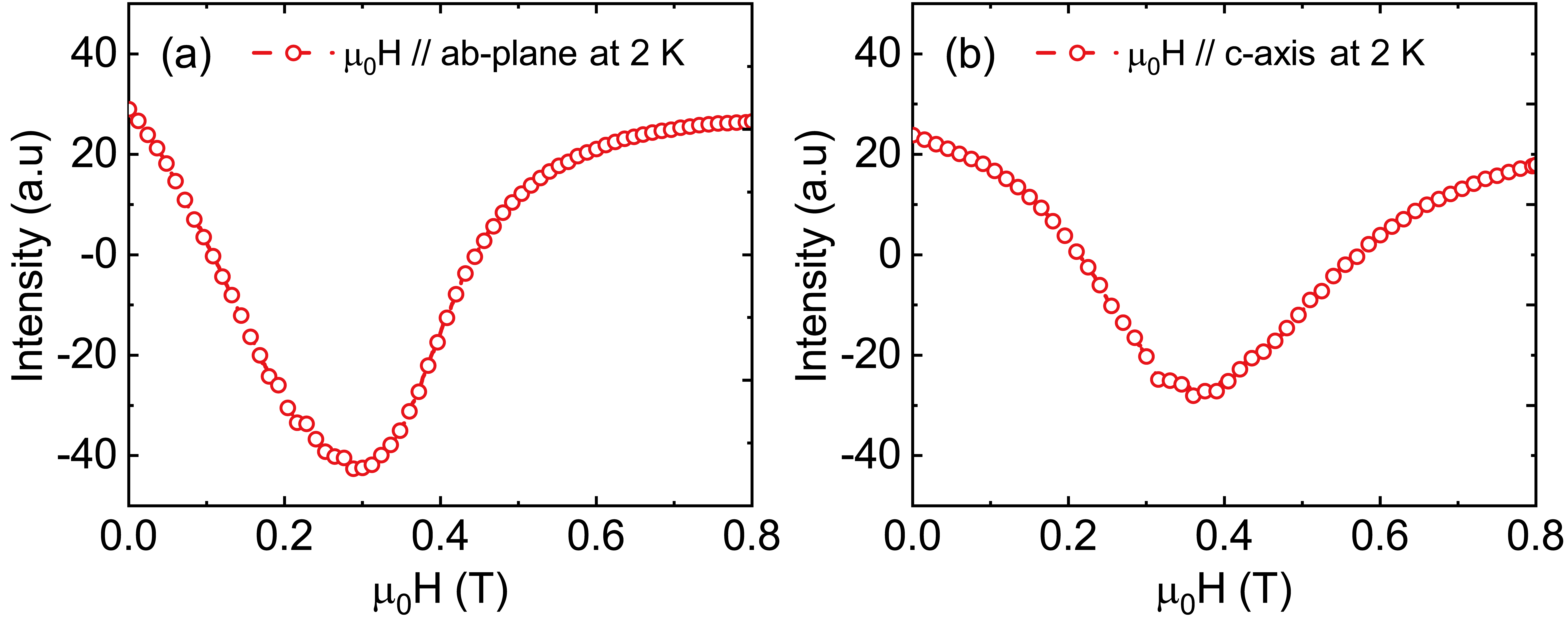}
	\caption{\label{fig:epsart}{(a) ESR spectrum in the ab-plane at 2 K (b) ESR spectrum along the c-axis at 2 K.}}
\end{figure}
\emph{ESR spectrum of \ce{KErTe2}}---We also conducted ESR spectroscopy measurements on \ce{KErTe2} both in the ab-plane and along the c-axis at a temperature of 2 K, as depicted in Fig. 6. 
The ESR spectra, whether in ab-plane or along c-axis, exhibit several broad linewidths.
The linewidth of ESR is related to the energy gap between the ground state and excited state of CEF.
The relationship between ESR linewidth and CEF energy gap is semi-quantitatively described by the following formula\cite{Sichelschmidt2020,Zhang2020}:
\begin{equation}
	{\mu _0}\Delta H \simeq \frac{1}{{\exp \left( {\Delta E/T} \right) - 1}}
\end{equation}
where $\mu_{0}$ is vacuum permeability, $\Delta H$ is the full width at half maximum (FWHM), $\Delta E$ is the energy gap from the ground state of CEF to the 1st excited state, and $T$ is the temperature. 

From this formula, we can see that when the CEF energy gap is smaller at a fixed temperature, the linewidth of the measured ESR is wider.
For some rare-earth chalcogenides with larger CEF energy gaps, such as \ce{NaYbS2}\cite{Sichelschmidt_2019} and \ce{NaYbSe2}\cite{PhysRevB.103.184419}, we can observe a narrow linewidth in the ESR spectra at low temperatures.
For \ce{KErTe2}, the CEF energy gap obtained from fitting the magnetization data is approximately 0.49 meV, which is much smaller than the gaps of \ce{NaYbS2} and \ce{NaYbSe2}. Therefore, we can observe a significantly broader linewidth in the ESR spectra of \ce{KErTe2} that exceeds the measurement range of the instrument.
Due to this broader linewidth, the X-band ($\sim$ 9.4 GHz) ESR spectrometer cannot observe a complete ESR spectrum, thereby making it difficult to accurately determine the Landé g-factors at low temperatures.
In the future, with high-frequency, high-field ESR spectrometers, we can analyze the CEF excitations and magnetic anisotropy of \ce{Er^{3+}} ions more accurately. 

\emph{Summary}---In our study, 
we derive the magnetic effective Hamiltonian to describe the magnetism of \ce{Er^{3+}} in \ce{KErTe2} based on the angular momentum $\hat{J}$ after SOC and symmetry analysis.
The effective Hamiltonian comprises terms of the CEF, interactions of the angular momentum $\hat{J}$, and the Zeeman splitting term.
By applying the mean-field approximation, we can effectively simplify the magnetic effective Hamiltonian that involves two-body operators.
The mean-field Hamiltonian is applicable for the analysis of the paramagnetic state at finite temperatures.
Therefore, we have successfully simulated the magnetization and heat capacity data of \ce{KErTe2} under different magnetic fields and obtained the CEF parameters and partial angular momentum $\hat{J}$ exchange interactions.
Simultaneously, using these fitted CEF parameters, we conducted calculations to determine the CEF energy levels and wave functions of \ce{KErTe2}. Our findings revealed a relatively small energy gap between the CEF ground state and the first excited state of \ce{KErTe2}. 
This feature can well explain a broad linewidth generated in the ESR spectra.
We also discussed the magnetic anisotropy in \ce{KErTe2}, where the \ce{Te^{2-}} anions would affect the magnetic anisotropy.
In sum, our utilization of the CEF theory and mean-field theory has enabled us to comprehensively explore and elucidate the magnetism of \ce{KErTe2} at finite temperatures.

\emph{Acknowledgments}---This work was supported by the National Science Foundation of China (Grant Nos. U1932215 and 12274186), the National Key Research and Development Program of China (Grant No. 2022YFA1402704), the Strategic Priority Research Program of the Chinese Academy of Sciences (Grant No. XDB33010100), and the Synergetic Extreme Condition User Facility (SECUF). A portion of this work was performed at the Steady High Magnetic Field Facilities, High Magnetic Field Laboratory, CAS.


\begin{thebibliography}{31}%
	\makeatletter
	\providecommand \@ifxundefined [1]{%
		\@ifx{#1\undefined}
	}%
	\providecommand \@ifnum [1]{%
		\ifnum #1\expandafter \@firstoftwo
		\else \expandafter \@secondoftwo
		\fi
	}%
	\providecommand \@ifx [1]{%
		\ifx #1\expandafter \@firstoftwo
		\else \expandafter \@secondoftwo
		\fi
	}%
	\providecommand \natexlab [1]{#1}%
	\providecommand \enquote  [1]{``#1''}%
	\providecommand \bibnamefont  [1]{#1}%
	\providecommand \bibfnamefont [1]{#1}%
	\providecommand \citenamefont [1]{#1}%
	\providecommand \href@noop [0]{\@secondoftwo}%
	\providecommand \href [0]{\begingroup \@sanitize@url \@href}%
	\providecommand \@href[1]{\@@startlink{#1}\@@href}%
	\providecommand \@@href[1]{\endgroup#1\@@endlink}%
	\providecommand \@sanitize@url [0]{\catcode `\\12\catcode `\$12\catcode
		`\&12\catcode `\#12\catcode `\^12\catcode `\_12\catcode `\%12\relax}%
	\providecommand \@@startlink[1]{}%
	\providecommand \@@endlink[0]{}%
	\providecommand \url  [0]{\begingroup\@sanitize@url \@url }%
	\providecommand \@url [1]{\endgroup\@href {#1}{\urlprefix }}%
	\providecommand \urlprefix  [0]{URL }%
	\providecommand \Eprint [0]{\href }%
	\providecommand \doibase [0]{http://dx.doi.org/}%
	\providecommand \selectlanguage [0]{\@gobble}%
	\providecommand \bibinfo  [0]{\@secondoftwo}%
	\providecommand \bibfield  [0]{\@secondoftwo}%
	\providecommand \translation [1]{[#1]}%
	\providecommand \BibitemOpen [0]{}%
	\providecommand \bibitemStop [0]{}%
	\providecommand \bibitemNoStop [0]{.\EOS\space}%
	\providecommand \EOS [0]{\spacefactor3000\relax}%
	\providecommand \BibitemShut  [1]{\csname bibitem#1\endcsname}%
	\let\auto@bib@innerbib\@empty
	\bibitem [{\citenamefont {Anderson}(1973)}]{Anderson1973}%
	\BibitemOpen
	\bibfield  {author} {\bibinfo {author} {\bibfnamefont {P.W.}\ \bibnamefont
			{Anderson}},\ }\bibfield  {title} {\enquote {\bibinfo {title} {Resonating
				valence bonds: A new kind of insulator?}}\ }\href {\doibase
		10.1016/0025-5408(73)90167-0} {\bibfield  {journal} {\bibinfo  {journal}
			{Mater. Res. Bull.}\ }\textbf {\bibinfo {volume} {8}},\ \bibinfo {pages}
		{153--160} (\bibinfo {year} {1973})}\BibitemShut {NoStop}%
	\bibitem [{\citenamefont {Anderson}(1987)}]{ANDERSON1987}%
	\BibitemOpen
	\bibfield  {author} {\bibinfo {author} {\bibfnamefont {P.W.}\ \bibnamefont
			{Anderson}},\ }\bibfield  {title} {\enquote {\bibinfo {title} {The resonating
				valence bond state in \ce{La2CuO4} and superconductivity},}\ }\href {\doibase
		10.1126/science.235.4793.1196} {\bibfield  {journal} {\bibinfo  {journal}
			{Science}\ }\textbf {\bibinfo {volume} {235}},\ \bibinfo {pages} {1196--1198}
		(\bibinfo {year} {1987})}\BibitemShut {NoStop}%
	\bibitem [{\citenamefont {Liu}\ \emph {et~al.}(2018)\citenamefont {Liu},
		\citenamefont {Zhang}, \citenamefont {Ji}, \citenamefont {Liu}, \citenamefont
		{Li}, \citenamefont {Wang}, \citenamefont {Lei}, \citenamefont {Chen},\ and\
		\citenamefont {Zhang}}]{liu2018rare}%
	\BibitemOpen
	\bibfield  {author} {\bibinfo {author} {\bibfnamefont {Weiwei}\ \bibnamefont
			{Liu}}, \bibinfo {author} {\bibfnamefont {Zheng}\ \bibnamefont {Zhang}},
		\bibinfo {author} {\bibfnamefont {Jianting}\ \bibnamefont {Ji}}, \bibinfo
		{author} {\bibfnamefont {Yixuan}\ \bibnamefont {Liu}}, \bibinfo {author}
		{\bibfnamefont {Jianshu}\ \bibnamefont {Li}}, \bibinfo {author}
		{\bibfnamefont {Xiaoqun}\ \bibnamefont {Wang}}, \bibinfo {author}
		{\bibfnamefont {Hechang}\ \bibnamefont {Lei}}, \bibinfo {author}
		{\bibfnamefont {Gang}\ \bibnamefont {Chen}}, \ and\ \bibinfo {author}
		{\bibfnamefont {Qingming}\ \bibnamefont {Zhang}},\ }\bibfield  {title}
	{\enquote {\bibinfo {title} {Rare-earth chalcogenides: A large family of
				triangular lattice spin liquid candidates},}\ }\href {\doibase
		10.1088/0256-307x/35/11/117501} {\bibfield  {journal} {\bibinfo  {journal}
			{Chin. Phys. Lett.}\ }\textbf {\bibinfo {volume} {35}},\ \bibinfo {pages}
		{117501} (\bibinfo {year} {2018})}\BibitemShut {NoStop}%
	\bibitem [{\citenamefont {Bordelon}\ \emph {et~al.}(2019)\citenamefont
		{Bordelon}, \citenamefont {Kenney}, \citenamefont {Liu}, \citenamefont
		{Hogan}, \citenamefont {Posthuma}, \citenamefont {Kavand}, \citenamefont
		{Lyu}, \citenamefont {Sherwin}, \citenamefont {Butch}, \citenamefont {Brown},
		\citenamefont {Graf}, \citenamefont {Balents},\ and\ \citenamefont
		{Wilson}}]{bordelon2019field}%
	\BibitemOpen
	\bibfield  {author} {\bibinfo {author} {\bibfnamefont {Mitchell~M.}\
			\bibnamefont {Bordelon}}, \bibinfo {author} {\bibfnamefont {Eric}\
			\bibnamefont {Kenney}}, \bibinfo {author} {\bibfnamefont {Chunxiao}\
			\bibnamefont {Liu}}, \bibinfo {author} {\bibfnamefont {Tom}\ \bibnamefont
			{Hogan}}, \bibinfo {author} {\bibfnamefont {Lorenzo}\ \bibnamefont
			{Posthuma}}, \bibinfo {author} {\bibfnamefont {Marzieh}\ \bibnamefont
			{Kavand}}, \bibinfo {author} {\bibfnamefont {Yuanqi}\ \bibnamefont {Lyu}},
		\bibinfo {author} {\bibfnamefont {Mark}\ \bibnamefont {Sherwin}}, \bibinfo
		{author} {\bibfnamefont {N.~P.}\ \bibnamefont {Butch}}, \bibinfo {author}
		{\bibfnamefont {Craig}\ \bibnamefont {Brown}}, \bibinfo {author}
		{\bibfnamefont {M.~J.}\ \bibnamefont {Graf}}, \bibinfo {author}
		{\bibfnamefont {Leon}\ \bibnamefont {Balents}}, \ and\ \bibinfo {author}
		{\bibfnamefont {Stephen~D.}\ \bibnamefont {Wilson}},\ }\bibfield  {title}
	{\enquote {\bibinfo {title} {Field-tunable quantum disordered ground state in
				the triangular-lattice antiferromagnet \ce{NaYbO2}},}\ }\href {\doibase
		10.1038/s41567-019-0594-5} {\bibfield  {journal} {\bibinfo  {journal} {Nat.
				Phys.}\ }\textbf {\bibinfo {volume} {15}},\ \bibinfo {pages} {1058--1064}
		(\bibinfo {year} {2019})}\BibitemShut {NoStop}%
	\bibitem [{\citenamefont {Ding}\ \emph {et~al.}(2019)\citenamefont {Ding},
		\citenamefont {Manuel}, \citenamefont {Bachus}, \citenamefont {Gru{\ss}ler},
		\citenamefont {Gegenwart}, \citenamefont {Singleton}, \citenamefont
		{Johnson}, \citenamefont {Walker}, \citenamefont {Adroja}, \citenamefont
		{Hillier},\ and\ \citenamefont {Tsirlin}}]{ding2019gapless}%
	\BibitemOpen
	\bibfield  {author} {\bibinfo {author} {\bibfnamefont {Lei}\ \bibnamefont
			{Ding}}, \bibinfo {author} {\bibfnamefont {Pascal}\ \bibnamefont {Manuel}},
		\bibinfo {author} {\bibfnamefont {Sebastian}\ \bibnamefont {Bachus}},
		\bibinfo {author} {\bibfnamefont {Franziska}\ \bibnamefont {Gru{\ss}ler}},
		\bibinfo {author} {\bibfnamefont {Philipp}\ \bibnamefont {Gegenwart}},
		\bibinfo {author} {\bibfnamefont {John}\ \bibnamefont {Singleton}}, \bibinfo
		{author} {\bibfnamefont {Roger~D.}\ \bibnamefont {Johnson}}, \bibinfo
		{author} {\bibfnamefont {Helen~C.}\ \bibnamefont {Walker}}, \bibinfo {author}
		{\bibfnamefont {Devashibhai~T.}\ \bibnamefont {Adroja}}, \bibinfo {author}
		{\bibfnamefont {Adrian~D.}\ \bibnamefont {Hillier}}, \ and\ \bibinfo {author}
		{\bibfnamefont {Alexander~A.}\ \bibnamefont {Tsirlin}},\ }\bibfield  {title}
	{\enquote {\bibinfo {title} {Gapless spin-liquid state in the structurally
				disorder-free triangular antiferromagnet \ce{NaYbO2}},}\ }\href {\doibase
		10.1103/physrevb.100.144432} {\bibfield  {journal} {\bibinfo  {journal}
			{Phys. Rev. B.}\ }\textbf {\bibinfo {volume} {100}},\ \bibinfo {pages}
		{144432} (\bibinfo {year} {2019})}\BibitemShut {NoStop}%
	\bibitem [{\citenamefont {Ranjith}\ \emph
		{et~al.}(2019{\natexlab{a}})\citenamefont {Ranjith}, \citenamefont
		{Dmytriieva}, \citenamefont {Khim}, \citenamefont {Sichelschmidt},
		\citenamefont {Luther}, \citenamefont {Ehlers}, \citenamefont {Yasuoka},
		\citenamefont {Wosnitza}, \citenamefont {Tsirlin}, \citenamefont {K\"uhne},\
		and\ \citenamefont {Baenitz}}]{ranjith2019field}%
	\BibitemOpen
	\bibfield  {author} {\bibinfo {author} {\bibfnamefont {K.~M.}\ \bibnamefont
			{Ranjith}}, \bibinfo {author} {\bibfnamefont {D.}~\bibnamefont {Dmytriieva}},
		\bibinfo {author} {\bibfnamefont {S.}~\bibnamefont {Khim}}, \bibinfo {author}
		{\bibfnamefont {J.}~\bibnamefont {Sichelschmidt}}, \bibinfo {author}
		{\bibfnamefont {S.}~\bibnamefont {Luther}}, \bibinfo {author} {\bibfnamefont
			{D.}~\bibnamefont {Ehlers}}, \bibinfo {author} {\bibfnamefont
			{H.}~\bibnamefont {Yasuoka}}, \bibinfo {author} {\bibfnamefont
			{J.}~\bibnamefont {Wosnitza}}, \bibinfo {author} {\bibfnamefont {A.~A.}\
			\bibnamefont {Tsirlin}}, \bibinfo {author} {\bibfnamefont {H.}~\bibnamefont
			{K\"uhne}}, \ and\ \bibinfo {author} {\bibfnamefont {M.}~\bibnamefont
			{Baenitz}},\ }\bibfield  {title} {\enquote {\bibinfo {title} {Field-induced
				instability of the quantum spin liquid ground state in the
				${J}_{\mathrm{eff}}=\frac{1}{2}$ triangular-lattice compound
				{${\mathrm{NaYbO}}_{2}$}},}\ }\href {\doibase 10.1103/PhysRevB.99.180401}
	{\bibfield  {journal} {\bibinfo  {journal} {Phys. Rev. B}\ }\textbf {\bibinfo
			{volume} {99}},\ \bibinfo {pages} {180401(R)} (\bibinfo {year}
		{2019}{\natexlab{a}})}\BibitemShut {NoStop}%
	\bibitem [{\citenamefont {Baenitz}\ \emph {et~al.}(2018)\citenamefont
		{Baenitz}, \citenamefont {Schlender}, \citenamefont {Sichelschmidt},
		\citenamefont {Onykiienko}, \citenamefont {Zangeneh}, \citenamefont
		{Ranjith}, \citenamefont {Sarkar}, \citenamefont {Hozoi}, \citenamefont
		{Walker}, \citenamefont {Orain}, \citenamefont {Yasuoka}, \citenamefont
		{van~den Brink}, \citenamefont {Klauss}, \citenamefont {Inosov},\ and\
		\citenamefont {Doert}}]{baenitz2018naybs}%
	\BibitemOpen
	\bibfield  {author} {\bibinfo {author} {\bibfnamefont {M.}~\bibnamefont
			{Baenitz}}, \bibinfo {author} {\bibfnamefont {Ph.}\ \bibnamefont
			{Schlender}}, \bibinfo {author} {\bibfnamefont {J.}~\bibnamefont
			{Sichelschmidt}}, \bibinfo {author} {\bibfnamefont {Y.~A.}\ \bibnamefont
			{Onykiienko}}, \bibinfo {author} {\bibfnamefont {Z.}~\bibnamefont
			{Zangeneh}}, \bibinfo {author} {\bibfnamefont {K.~M.}\ \bibnamefont
			{Ranjith}}, \bibinfo {author} {\bibfnamefont {R.}~\bibnamefont {Sarkar}},
		\bibinfo {author} {\bibfnamefont {L.}~\bibnamefont {Hozoi}}, \bibinfo
		{author} {\bibfnamefont {H.~C.}\ \bibnamefont {Walker}}, \bibinfo {author}
		{\bibfnamefont {J.-C.}\ \bibnamefont {Orain}}, \bibinfo {author}
		{\bibfnamefont {H.}~\bibnamefont {Yasuoka}}, \bibinfo {author} {\bibfnamefont
			{J.}~\bibnamefont {van~den Brink}}, \bibinfo {author} {\bibfnamefont {H.~H.}\
			\bibnamefont {Klauss}}, \bibinfo {author} {\bibfnamefont {D.~S.}\
			\bibnamefont {Inosov}}, \ and\ \bibinfo {author} {\bibfnamefont {Th.}\
			\bibnamefont {Doert}},\ }\bibfield  {title} {\enquote {\bibinfo {title}
			{\ce{NaYbS2}: A planar spin-1/2 triangular-lattice magnet and putative spin
				liquid},}\ }\href {\doibase 10.1103/physrevb.98.220409} {\bibfield  {journal}
		{\bibinfo  {journal} {Phys. Rev. B.}\ }\textbf {\bibinfo {volume} {98}},\
		\bibinfo {pages} {220409(R)} (\bibinfo {year} {2018})}\BibitemShut {NoStop}%
	\bibitem [{\citenamefont {Sarkar}\ \emph {et~al.}(2019)\citenamefont {Sarkar},
		\citenamefont {Schlender}, \citenamefont {Grinenko}, \citenamefont
		{Haeussler}, \citenamefont {Baker}, \citenamefont {Doert},\ and\
		\citenamefont {Klauss}}]{sarkar2019quantum}%
	\BibitemOpen
	\bibfield  {author} {\bibinfo {author} {\bibfnamefont {R.}~\bibnamefont
			{Sarkar}}, \bibinfo {author} {\bibfnamefont {Ph.}\ \bibnamefont {Schlender}},
		\bibinfo {author} {\bibfnamefont {V.}~\bibnamefont {Grinenko}}, \bibinfo
		{author} {\bibfnamefont {E.}~\bibnamefont {Haeussler}}, \bibinfo {author}
		{\bibfnamefont {Peter~J.}\ \bibnamefont {Baker}}, \bibinfo {author}
		{\bibfnamefont {Th.}\ \bibnamefont {Doert}}, \ and\ \bibinfo {author}
		{\bibfnamefont {H.-H.}\ \bibnamefont {Klauss}},\ }\bibfield  {title}
	{\enquote {\bibinfo {title} {Quantum spin liquid ground state in the disorder
				free triangular lattice \ce{NaYbS2}},}\ }\href {\doibase
		10.1103/physrevb.100.241116} {\bibfield  {journal} {\bibinfo  {journal}
			{Phys. Rev. B.}\ }\textbf {\bibinfo {volume} {100}},\ \bibinfo {pages}
		{241116(R)} (\bibinfo {year} {2019})}\BibitemShut {NoStop}%
	\bibitem [{\citenamefont {Zhang}\ \emph
		{et~al.}(2021{\natexlab{a}})\citenamefont {Zhang}, \citenamefont {Ma},
		\citenamefont {Li}, \citenamefont {Wang}, \citenamefont {Adroja},
		\citenamefont {Perring}, \citenamefont {Liu}, \citenamefont {Jin},
		\citenamefont {Ji}, \citenamefont {Wang}, \citenamefont {Kamiya},
		\citenamefont {Wang}, \citenamefont {Ma},\ and\ \citenamefont
		{Zhang}}]{Zhang2020}%
	\BibitemOpen
	\bibfield  {author} {\bibinfo {author} {\bibfnamefont {Zheng}\ \bibnamefont
			{Zhang}}, \bibinfo {author} {\bibfnamefont {Xiaoli}\ \bibnamefont {Ma}},
		\bibinfo {author} {\bibfnamefont {Jianshu}\ \bibnamefont {Li}}, \bibinfo
		{author} {\bibfnamefont {Guohua}\ \bibnamefont {Wang}}, \bibinfo {author}
		{\bibfnamefont {D.~T.}\ \bibnamefont {Adroja}}, \bibinfo {author}
		{\bibfnamefont {T.~P.}\ \bibnamefont {Perring}}, \bibinfo {author}
		{\bibfnamefont {Weiwei}\ \bibnamefont {Liu}}, \bibinfo {author}
		{\bibfnamefont {Feng}\ \bibnamefont {Jin}}, \bibinfo {author} {\bibfnamefont
			{Jianting}\ \bibnamefont {Ji}}, \bibinfo {author} {\bibfnamefont {Yimeng}\
			\bibnamefont {Wang}}, \bibinfo {author} {\bibfnamefont {Yoshitomo}\
			\bibnamefont {Kamiya}}, \bibinfo {author} {\bibfnamefont {Xiaoqun}\
			\bibnamefont {Wang}}, \bibinfo {author} {\bibfnamefont {Jie}\ \bibnamefont
			{Ma}}, \ and\ \bibinfo {author} {\bibfnamefont {Qingming}\ \bibnamefont
			{Zhang}},\ }\bibfield  {title} {\enquote {\bibinfo {title} {Crystalline
				electric field excitations in the quantum spin liquid candidate
				\ce{NaYbSe2}},}\ }\href {\doibase 10.1103/PhysRevB.103.035144} {\bibfield
		{journal} {\bibinfo  {journal} {Phys. Rev. B}\ }\textbf {\bibinfo {volume}
			{103}},\ \bibinfo {pages} {035144} (\bibinfo {year}
		{2021}{\natexlab{a}})}\BibitemShut {NoStop}%
	\bibitem [{\citenamefont {Ranjith}\ \emph
		{et~al.}(2019{\natexlab{b}})\citenamefont {Ranjith}, \citenamefont {Luther},
		\citenamefont {Reimann}, \citenamefont {Schmidt}, \citenamefont {Schlender},
		\citenamefont {Sichelschmidt}, \citenamefont {Yasuoka}, \citenamefont
		{Strydom}, \citenamefont {Skourski}, \citenamefont {Wosnitza}, \citenamefont
		{K\"uhne}, \citenamefont {Doert},\ and\ \citenamefont
		{Baenitz}}]{PhysRevB.100.224417}%
	\BibitemOpen
	\bibfield  {author} {\bibinfo {author} {\bibfnamefont {K.~M.}\ \bibnamefont
			{Ranjith}}, \bibinfo {author} {\bibfnamefont {S.}~\bibnamefont {Luther}},
		\bibinfo {author} {\bibfnamefont {T.}~\bibnamefont {Reimann}}, \bibinfo
		{author} {\bibfnamefont {B.}~\bibnamefont {Schmidt}}, \bibinfo {author}
		{\bibfnamefont {Ph.}\ \bibnamefont {Schlender}}, \bibinfo {author}
		{\bibfnamefont {J.}~\bibnamefont {Sichelschmidt}}, \bibinfo {author}
		{\bibfnamefont {H.}~\bibnamefont {Yasuoka}}, \bibinfo {author} {\bibfnamefont
			{A.~M.}\ \bibnamefont {Strydom}}, \bibinfo {author} {\bibfnamefont
			{Y.}~\bibnamefont {Skourski}}, \bibinfo {author} {\bibfnamefont
			{J.}~\bibnamefont {Wosnitza}}, \bibinfo {author} {\bibfnamefont
			{H.}~\bibnamefont {K\"uhne}}, \bibinfo {author} {\bibfnamefont {Th.}\
			\bibnamefont {Doert}}, \ and\ \bibinfo {author} {\bibfnamefont
			{M.}~\bibnamefont {Baenitz}},\ }\bibfield  {title} {\enquote {\bibinfo
			{title} {Anisotropic field-induced ordering in the triangular-lattice quantum
				spin liquid \ce{NaYbSe2}},}\ }\href {\doibase 10.1103/PhysRevB.100.224417}
	{\bibfield  {journal} {\bibinfo  {journal} {Phys. Rev. B}\ }\textbf {\bibinfo
			{volume} {100}},\ \bibinfo {pages} {224417} (\bibinfo {year}
		{2019}{\natexlab{b}})}\BibitemShut {NoStop}%
	\bibitem [{\citenamefont {Helton}\ \emph {et~al.}(2007)\citenamefont {Helton},
		\citenamefont {Matan}, \citenamefont {Shores}, \citenamefont {Nytko},
		\citenamefont {Bartlett}, \citenamefont {Yoshida}, \citenamefont {Takano},
		\citenamefont {Suslov}, \citenamefont {Qiu}, \citenamefont {Chung},
		\citenamefont {Nocera},\ and\ \citenamefont {Lee}}]{PhysRevLett.98.107204}%
	\BibitemOpen
	\bibfield  {author} {\bibinfo {author} {\bibfnamefont {J.~S.}\ \bibnamefont
			{Helton}}, \bibinfo {author} {\bibfnamefont {K.}~\bibnamefont {Matan}},
		\bibinfo {author} {\bibfnamefont {M.~P.}\ \bibnamefont {Shores}}, \bibinfo
		{author} {\bibfnamefont {E.~A.}\ \bibnamefont {Nytko}}, \bibinfo {author}
		{\bibfnamefont {B.~M.}\ \bibnamefont {Bartlett}}, \bibinfo {author}
		{\bibfnamefont {Y.}~\bibnamefont {Yoshida}}, \bibinfo {author} {\bibfnamefont
			{Y.}~\bibnamefont {Takano}}, \bibinfo {author} {\bibfnamefont
			{A.}~\bibnamefont {Suslov}}, \bibinfo {author} {\bibfnamefont
			{Y.}~\bibnamefont {Qiu}}, \bibinfo {author} {\bibfnamefont {J.-H.}\
			\bibnamefont {Chung}}, \bibinfo {author} {\bibfnamefont {D.~G.}\ \bibnamefont
			{Nocera}}, \ and\ \bibinfo {author} {\bibfnamefont {Y.~S.}\ \bibnamefont
			{Lee}},\ }\bibfield  {title} {\enquote {\bibinfo {title} {Spin dynamics of
				the spin-$1/2$ kagome lattice antiferromagnet
				{${\mathrm{ZnCu}}_{3}(\mathrm{OH}{)}_{6}{\mathrm{Cl}}_{2}$}},}\ }\href
	{\doibase 10.1103/PhysRevLett.98.107204} {\bibfield  {journal} {\bibinfo
			{journal} {Phys. Rev. Lett.}\ }\textbf {\bibinfo {volume} {98}},\ \bibinfo
		{pages} {107204} (\bibinfo {year} {2007})}\BibitemShut {NoStop}%
	\bibitem [{\citenamefont {Itou}\ \emph {et~al.}(2008)\citenamefont {Itou},
		\citenamefont {Oyamada}, \citenamefont {Maegawa}, \citenamefont {Tamura},\
		and\ \citenamefont {Kato}}]{itou2008quantum}%
	\BibitemOpen
	\bibfield  {author} {\bibinfo {author} {\bibfnamefont {T.}~\bibnamefont
			{Itou}}, \bibinfo {author} {\bibfnamefont {A.}~\bibnamefont {Oyamada}},
		\bibinfo {author} {\bibfnamefont {S.}~\bibnamefont {Maegawa}}, \bibinfo
		{author} {\bibfnamefont {M.}~\bibnamefont {Tamura}}, \ and\ \bibinfo {author}
		{\bibfnamefont {R.}~\bibnamefont {Kato}},\ }\bibfield  {title} {\enquote
		{\bibinfo {title} {Quantum spin liquid in the spin-1/ 2 triangular
				antiferromagnet \ce{EtMe3Sb[Pd(dmit)2]2}},}\ }\href {\doibase
		10.1103/physrevb.77.104413} {\bibfield  {journal} {\bibinfo  {journal} {Phys.
				Rev. B.}\ }\textbf {\bibinfo {volume} {77}},\ \bibinfo {pages} {104413}
		(\bibinfo {year} {2008})}\BibitemShut {NoStop}%
	\bibitem [{\citenamefont {Li}\ \emph {et~al.}(2015{\natexlab{a}})\citenamefont
		{Li}, \citenamefont {Liao}, \citenamefont {Zhang}, \citenamefont {Li},
		\citenamefont {Jin}, \citenamefont {Ling}, \citenamefont {Zhang},
		\citenamefont {Zou}, \citenamefont {Pi}, \citenamefont {Yang}, \citenamefont
		{Wang}, \citenamefont {Wu},\ and\ \citenamefont {Zhang}}]{li2015gapless}%
	\BibitemOpen
	\bibfield  {author} {\bibinfo {author} {\bibfnamefont {Yuesheng}\
			\bibnamefont {Li}}, \bibinfo {author} {\bibfnamefont {Haijun}\ \bibnamefont
			{Liao}}, \bibinfo {author} {\bibfnamefont {Zhen}\ \bibnamefont {Zhang}},
		\bibinfo {author} {\bibfnamefont {Shiyan}\ \bibnamefont {Li}}, \bibinfo
		{author} {\bibfnamefont {Feng}\ \bibnamefont {Jin}}, \bibinfo {author}
		{\bibfnamefont {Langsheng}\ \bibnamefont {Ling}}, \bibinfo {author}
		{\bibfnamefont {Lei}\ \bibnamefont {Zhang}}, \bibinfo {author} {\bibfnamefont
			{Youming}\ \bibnamefont {Zou}}, \bibinfo {author} {\bibfnamefont
			{Li}~\bibnamefont {Pi}}, \bibinfo {author} {\bibfnamefont {Zhaorong}\
			\bibnamefont {Yang}}, \bibinfo {author} {\bibfnamefont {Junfeng}\
			\bibnamefont {Wang}}, \bibinfo {author} {\bibfnamefont {Zhonghua}\
			\bibnamefont {Wu}}, \ and\ \bibinfo {author} {\bibfnamefont {Qingming}\
			\bibnamefont {Zhang}},\ }\bibfield  {title} {\enquote {\bibinfo {title}
			{Gapless quantum spin liquid ground state in the two-dimensional spin-1/2
				triangular antiferromagnet \ce{YbMgGaO4}},}\ }\href {\doibase
		10.1038/srep16419} {\bibfield  {journal} {\bibinfo  {journal} {Sci. Rep.}\
		}\textbf {\bibinfo {volume} {5}},\ \bibinfo {pages} {16419} (\bibinfo {year}
		{2015}{\natexlab{a}})}\BibitemShut {NoStop}%
	\bibitem [{\citenamefont {Li}\ \emph {et~al.}(2015{\natexlab{b}})\citenamefont
		{Li}, \citenamefont {Chen}, \citenamefont {Tong}, \citenamefont {Pi},
		\citenamefont {Liu}, \citenamefont {Yang}, \citenamefont {Wang},\ and\
		\citenamefont {Zhang}}]{li2015rare}%
	\BibitemOpen
	\bibfield  {author} {\bibinfo {author} {\bibfnamefont {Yuesheng}\
			\bibnamefont {Li}}, \bibinfo {author} {\bibfnamefont {Gang}\ \bibnamefont
			{Chen}}, \bibinfo {author} {\bibfnamefont {Wei}\ \bibnamefont {Tong}},
		\bibinfo {author} {\bibfnamefont {Li}~\bibnamefont {Pi}}, \bibinfo {author}
		{\bibfnamefont {Juanjuan}\ \bibnamefont {Liu}}, \bibinfo {author}
		{\bibfnamefont {Zhaorong}\ \bibnamefont {Yang}}, \bibinfo {author}
		{\bibfnamefont {Xiaoqun}\ \bibnamefont {Wang}}, \ and\ \bibinfo {author}
		{\bibfnamefont {Qingming}\ \bibnamefont {Zhang}},\ }\bibfield  {title}
	{\enquote {\bibinfo {title} {Rare-earth triangular lattice spin liquid: a
				single-crystal study of \ce{YbMgGaO4}},}\ }\href {\doibase
		10.1103/physrevlett.115.167203} {\bibfield  {journal} {\bibinfo  {journal}
			{Phys. Rev. Lett.}\ }\textbf {\bibinfo {volume} {115}},\ \bibinfo {pages}
		{167203} (\bibinfo {year} {2015}{\natexlab{b}})}\BibitemShut {NoStop}%
	\bibitem [{\citenamefont {Li}\ \emph {et~al.}(2016{\natexlab{a}})\citenamefont
		{Li}, \citenamefont {Adroja}, \citenamefont {Biswas}, \citenamefont {Baker},
		\citenamefont {Zhang}, \citenamefont {Liu}, \citenamefont {Tsirlin},
		\citenamefont {Gegenwart},\ and\ \citenamefont
		{Zhang}}]{PhysRevLett.117.097201}%
	\BibitemOpen
	\bibfield  {author} {\bibinfo {author} {\bibfnamefont {Yuesheng}\
			\bibnamefont {Li}}, \bibinfo {author} {\bibfnamefont {Devashibhai}\
			\bibnamefont {Adroja}}, \bibinfo {author} {\bibfnamefont {Pabitra~K.}\
			\bibnamefont {Biswas}}, \bibinfo {author} {\bibfnamefont {Peter~J.}\
			\bibnamefont {Baker}}, \bibinfo {author} {\bibfnamefont {Qian}\ \bibnamefont
			{Zhang}}, \bibinfo {author} {\bibfnamefont {Juanjuan}\ \bibnamefont {Liu}},
		\bibinfo {author} {\bibfnamefont {Alexander~A.}\ \bibnamefont {Tsirlin}},
		\bibinfo {author} {\bibfnamefont {Philipp}\ \bibnamefont {Gegenwart}}, \ and\
		\bibinfo {author} {\bibfnamefont {Qingming}\ \bibnamefont {Zhang}},\
	}\bibfield  {title} {\enquote {\bibinfo {title} {Muon spin relaxation
				evidence for the {U}(1) quantum spin-liquid ground state in the triangular
				antiferromagnet {${\mathrm{YbMgGaO}}_{4}$}},}\ }\href {\doibase
		10.1103/PhysRevLett.117.097201} {\bibfield  {journal} {\bibinfo  {journal}
			{Phys. Rev. Lett.}\ }\textbf {\bibinfo {volume} {117}},\ \bibinfo {pages}
		{097201} (\bibinfo {year} {2016}{\natexlab{a}})}\BibitemShut {NoStop}%
	\bibitem [{\citenamefont {Xu}\ \emph {et~al.}(2016)\citenamefont {Xu},
		\citenamefont {Zhang}, \citenamefont {Li}, \citenamefont {Yu}, \citenamefont
		{Hong}, \citenamefont {Zhang},\ and\ \citenamefont
		{Li}}]{PhysRevLett.117.267202}%
	\BibitemOpen
	\bibfield  {author} {\bibinfo {author} {\bibfnamefont {Y.}~\bibnamefont
			{Xu}}, \bibinfo {author} {\bibfnamefont {J.}~\bibnamefont {Zhang}}, \bibinfo
		{author} {\bibfnamefont {Y.~S.}\ \bibnamefont {Li}}, \bibinfo {author}
		{\bibfnamefont {Y.~J.}\ \bibnamefont {Yu}}, \bibinfo {author} {\bibfnamefont
			{X.~C.}\ \bibnamefont {Hong}}, \bibinfo {author} {\bibfnamefont {Q.~M.}\
			\bibnamefont {Zhang}}, \ and\ \bibinfo {author} {\bibfnamefont {S.~Y.}\
			\bibnamefont {Li}},\ }\bibfield  {title} {\enquote {\bibinfo {title} {Absence
				of magnetic thermal conductivity in the quantum spin-liquid candidate
				{${\mathrm{YbMgGaO}}_{4}$}},}\ }\href {\doibase
		10.1103/PhysRevLett.117.267202} {\bibfield  {journal} {\bibinfo  {journal}
			{Phys. Rev. Lett.}\ }\textbf {\bibinfo {volume} {117}},\ \bibinfo {pages}
		{267202} (\bibinfo {year} {2016})}\BibitemShut {NoStop}%
	\bibitem [{\citenamefont {Shen}\ \emph {et~al.}(2016)\citenamefont {Shen},
		\citenamefont {Li}, \citenamefont {Wo}, \citenamefont {Li}, \citenamefont
		{Shen}, \citenamefont {Pan}, \citenamefont {Wang}, \citenamefont {Walker},
		\citenamefont {Steffens}, \citenamefont {Boehm}, \citenamefont {Hao},
		\citenamefont {Quintero-Castro}, \citenamefont {Harriger}, \citenamefont
		{Frontzek}, \citenamefont {Hao}, \citenamefont {Meng}, \citenamefont {Zhang},
		\citenamefont {Chen},\ and\ \citenamefont {Zhao}}]{Shen2016}%
	\BibitemOpen
	\bibfield  {author} {\bibinfo {author} {\bibfnamefont {Yao}\ \bibnamefont
			{Shen}}, \bibinfo {author} {\bibfnamefont {Yao-Dong}\ \bibnamefont {Li}},
		\bibinfo {author} {\bibfnamefont {Hongliang}\ \bibnamefont {Wo}}, \bibinfo
		{author} {\bibfnamefont {Yuesheng}\ \bibnamefont {Li}}, \bibinfo {author}
		{\bibfnamefont {Shoudong}\ \bibnamefont {Shen}}, \bibinfo {author}
		{\bibfnamefont {Bingying}\ \bibnamefont {Pan}}, \bibinfo {author}
		{\bibfnamefont {Qisi}\ \bibnamefont {Wang}}, \bibinfo {author} {\bibfnamefont
			{H.~C.}\ \bibnamefont {Walker}}, \bibinfo {author} {\bibfnamefont
			{P.}~\bibnamefont {Steffens}}, \bibinfo {author} {\bibfnamefont
			{M.}~\bibnamefont {Boehm}}, \bibinfo {author} {\bibfnamefont {Yiqing}\
			\bibnamefont {Hao}}, \bibinfo {author} {\bibfnamefont {D.~L.}\ \bibnamefont
			{Quintero-Castro}}, \bibinfo {author} {\bibfnamefont {L.~W.}\ \bibnamefont
			{Harriger}}, \bibinfo {author} {\bibfnamefont {M.~D.}\ \bibnamefont
			{Frontzek}}, \bibinfo {author} {\bibfnamefont {Lijie}\ \bibnamefont {Hao}},
		\bibinfo {author} {\bibfnamefont {Siqin}\ \bibnamefont {Meng}}, \bibinfo
		{author} {\bibfnamefont {Qingming}\ \bibnamefont {Zhang}}, \bibinfo {author}
		{\bibfnamefont {Gang}\ \bibnamefont {Chen}}, \ and\ \bibinfo {author}
		{\bibfnamefont {Jun}\ \bibnamefont {Zhao}},\ }\bibfield  {title} {\enquote
		{\bibinfo {title} {Evidence for a spinon fermi surface in a
				triangular-lattice quantum-spin-liquid candidate},}\ }\href {\doibase
		10.1038/nature20614} {\bibfield  {journal} {\bibinfo  {journal} {Nature}\
		}\textbf {\bibinfo {volume} {540}},\ \bibinfo {pages} {559--562} (\bibinfo
		{year} {2016})}\BibitemShut {NoStop}%
	\bibitem [{\citenamefont {Paddison}\ \emph {et~al.}(2016)\citenamefont
		{Paddison}, \citenamefont {Daum}, \citenamefont {Dun}, \citenamefont
		{Ehlers}, \citenamefont {Liu}, \citenamefont {Stone}, \citenamefont {Zhou},\
		and\ \citenamefont {Mourigal}}]{Paddison2016}%
	\BibitemOpen
	\bibfield  {author} {\bibinfo {author} {\bibfnamefont {Joseph A.~M.}\
			\bibnamefont {Paddison}}, \bibinfo {author} {\bibfnamefont {Marcus}\
			\bibnamefont {Daum}}, \bibinfo {author} {\bibfnamefont {Zhiling}\
			\bibnamefont {Dun}}, \bibinfo {author} {\bibfnamefont {Georg}\ \bibnamefont
			{Ehlers}}, \bibinfo {author} {\bibfnamefont {Yaohua}\ \bibnamefont {Liu}},
		\bibinfo {author} {\bibfnamefont {Matthew~B.}\ \bibnamefont {Stone}},
		\bibinfo {author} {\bibfnamefont {Haidong}\ \bibnamefont {Zhou}}, \ and\
		\bibinfo {author} {\bibfnamefont {Martin}\ \bibnamefont {Mourigal}},\
	}\bibfield  {title} {\enquote {\bibinfo {title} {Continuous excitations of
				the triangular-lattice quantum spin liquid \ce{YbMgGaO4}},}\ }\href {\doibase
		10.1038/nphys3971} {\bibfield  {journal} {\bibinfo  {journal} {Nat. Phys.}\
		}\textbf {\bibinfo {volume} {13}},\ \bibinfo {pages} {117--122} (\bibinfo
		{year} {2016})}\BibitemShut {NoStop}%
	\bibitem [{\citenamefont {Xing}\ \emph {et~al.}(2019)\citenamefont {Xing},
		\citenamefont {Sanjeewa}, \citenamefont {Kim}, \citenamefont {Meier},
		\citenamefont {May}, \citenamefont {Zheng}, \citenamefont {Custelcean},
		\citenamefont {Stewart},\ and\ \citenamefont
		{Sefat}}]{PhysRevMaterials.3.114413}%
	\BibitemOpen
	\bibfield  {author} {\bibinfo {author} {\bibfnamefont {Jie}\ \bibnamefont
			{Xing}}, \bibinfo {author} {\bibfnamefont {Liurukara~D.}\ \bibnamefont
			{Sanjeewa}}, \bibinfo {author} {\bibfnamefont {Jungsoo}\ \bibnamefont {Kim}},
		\bibinfo {author} {\bibfnamefont {William~R.}\ \bibnamefont {Meier}},
		\bibinfo {author} {\bibfnamefont {Andrew~F.}\ \bibnamefont {May}}, \bibinfo
		{author} {\bibfnamefont {Qiang}\ \bibnamefont {Zheng}}, \bibinfo {author}
		{\bibfnamefont {Radu}\ \bibnamefont {Custelcean}}, \bibinfo {author}
		{\bibfnamefont {G.~R.}\ \bibnamefont {Stewart}}, \ and\ \bibinfo {author}
		{\bibfnamefont {Athena~S.}\ \bibnamefont {Sefat}},\ }\bibfield  {title}
	{\enquote {\bibinfo {title} {Synthesis, magnetization, and heat capacity of
				triangular lattice materials {${\mathrm{NaErSe}}_{2}$} and
				{${\mathrm{KErSe}}_{2}$}},}\ }\href {\doibase
		10.1103/PhysRevMaterials.3.114413} {\bibfield  {journal} {\bibinfo  {journal}
			{Phys. Rev. Materials}\ }\textbf {\bibinfo {volume} {3}},\ \bibinfo {pages}
		{114413} (\bibinfo {year} {2019})}\BibitemShut {NoStop}%
	\bibitem [{\citenamefont {Scheie}\ \emph {et~al.}(2020)\citenamefont {Scheie},
		\citenamefont {Garlea}, \citenamefont {Sanjeewa}, \citenamefont {Xing},\ and\
		\citenamefont {Sefat}}]{PhysRevB.101.144432}%
	\BibitemOpen
	\bibfield  {author} {\bibinfo {author} {\bibfnamefont {A.}~\bibnamefont
			{Scheie}}, \bibinfo {author} {\bibfnamefont {V.~O.}\ \bibnamefont {Garlea}},
		\bibinfo {author} {\bibfnamefont {L.~D.}\ \bibnamefont {Sanjeewa}}, \bibinfo
		{author} {\bibfnamefont {J.}~\bibnamefont {Xing}}, \ and\ \bibinfo {author}
		{\bibfnamefont {A.~S.}\ \bibnamefont {Sefat}},\ }\bibfield  {title} {\enquote
		{\bibinfo {title} {Crystal-field hamiltonian and anisotropy in
				{${\mathrm{KErSe}}_{2}$} and {${\mathrm{CsErSe}}_{2}$}},}\ }\href {\doibase
		10.1103/PhysRevB.101.144432} {\bibfield  {journal} {\bibinfo  {journal}
			{Phys. Rev. B}\ }\textbf {\bibinfo {volume} {101}},\ \bibinfo {pages}
		{144432} (\bibinfo {year} {2020})}\BibitemShut {NoStop}%
	\bibitem [{\citenamefont {Zhang}\ \emph
		{et~al.}(2021{\natexlab{b}})\citenamefont {Zhang}, \citenamefont {Li},
		\citenamefont {Liu}, \citenamefont {Zhang}, \citenamefont {Ji}, \citenamefont
		{Jin}, \citenamefont {Chen}, \citenamefont {Wang}, \citenamefont {Wang},
		\citenamefont {Ma},\ and\ \citenamefont {Zhang}}]{PhysRevB.103.184419}%
	\BibitemOpen
	\bibfield  {author} {\bibinfo {author} {\bibfnamefont {Zheng}\ \bibnamefont
			{Zhang}}, \bibinfo {author} {\bibfnamefont {Jianshu}\ \bibnamefont {Li}},
		\bibinfo {author} {\bibfnamefont {Weiwei}\ \bibnamefont {Liu}}, \bibinfo
		{author} {\bibfnamefont {Zhitao}\ \bibnamefont {Zhang}}, \bibinfo {author}
		{\bibfnamefont {Jianting}\ \bibnamefont {Ji}}, \bibinfo {author}
		{\bibfnamefont {Feng}\ \bibnamefont {Jin}}, \bibinfo {author} {\bibfnamefont
			{Rui}\ \bibnamefont {Chen}}, \bibinfo {author} {\bibfnamefont {Junfeng}\
			\bibnamefont {Wang}}, \bibinfo {author} {\bibfnamefont {Xiaoqun}\
			\bibnamefont {Wang}}, \bibinfo {author} {\bibfnamefont {Jie}\ \bibnamefont
			{Ma}}, \ and\ \bibinfo {author} {\bibfnamefont {Qingming}\ \bibnamefont
			{Zhang}},\ }\bibfield  {title} {\enquote {\bibinfo {title} {Effective
				magnetic hamiltonian at finite temperatures for rare-earth chalcogenides},}\
	}\href {\doibase 10.1103/PhysRevB.103.184419} {\bibfield  {journal} {\bibinfo
			{journal} {Phys. Rev. B}\ }\textbf {\bibinfo {volume} {103}},\ \bibinfo
		{pages} {184419} (\bibinfo {year} {2021}{\natexlab{b}})}\BibitemShut
	{NoStop}%
	\bibitem [{\citenamefont {Liu}\ \emph {et~al.}(2021)\citenamefont {Liu},
		\citenamefont {Yan}, \citenamefont {Zhang}, \citenamefont {Ji}, \citenamefont
		{Shi}, \citenamefont {Jin},\ and\ \citenamefont {Zhang}}]{liu2021}%
	\BibitemOpen
	\bibfield  {author} {\bibinfo {author} {\bibfnamefont {Weiwei}\ \bibnamefont
			{Liu}}, \bibinfo {author} {\bibfnamefont {Dayu}\ \bibnamefont {Yan}},
		\bibinfo {author} {\bibfnamefont {Zheng}\ \bibnamefont {Zhang}}, \bibinfo
		{author} {\bibfnamefont {Jianting}\ \bibnamefont {Ji}}, \bibinfo {author}
		{\bibfnamefont {Youguo}\ \bibnamefont {Shi}}, \bibinfo {author}
		{\bibfnamefont {Feng}\ \bibnamefont {Jin}}, \ and\ \bibinfo {author}
		{\bibfnamefont {Qingming}\ \bibnamefont {Zhang}},\ }\bibfield  {title}
	{\enquote {\bibinfo {title} {Crystal growth and magnetic properties of
				quantum spin liquid candidate \ce{KErTe2}},}\ }\href {\doibase
		10.1088/1674-1056/ac1574} {\bibfield  {journal} {\bibinfo  {journal} {Chinese
				Physics B}\ }\textbf {\bibinfo {volume} {30}},\ \bibinfo {pages} {107504}
		(\bibinfo {year} {2021})}\BibitemShut {NoStop}%
	\bibitem [{\citenamefont {Zhang}\ \emph {et~al.}(2020)\citenamefont {Zhang},
		\citenamefont {Yin}, \citenamefont {Ma}, \citenamefont {Liu}, \citenamefont
		{Li}, \citenamefont {Jin}, \citenamefont {Ji}, \citenamefont {Wang},
		\citenamefont {Wang}, \citenamefont {Yu},\ and\ \citenamefont
		{Zhang}}]{zhang2020pressure}%
	\BibitemOpen
	\bibfield  {author} {\bibinfo {author} {\bibfnamefont {Zheng}\ \bibnamefont
			{Zhang}}, \bibinfo {author} {\bibfnamefont {Yunyu}\ \bibnamefont {Yin}},
		\bibinfo {author} {\bibfnamefont {Xiaoli}\ \bibnamefont {Ma}}, \bibinfo
		{author} {\bibfnamefont {Weiwei}\ \bibnamefont {Liu}}, \bibinfo {author}
		{\bibfnamefont {Jianshu}\ \bibnamefont {Li}}, \bibinfo {author}
		{\bibfnamefont {Feng}\ \bibnamefont {Jin}}, \bibinfo {author} {\bibfnamefont
			{Jianting}\ \bibnamefont {Ji}}, \bibinfo {author} {\bibfnamefont {Yimeng}\
			\bibnamefont {Wang}}, \bibinfo {author} {\bibfnamefont {Xiaoqun}\
			\bibnamefont {Wang}}, \bibinfo {author} {\bibfnamefont {Xiaohui}\
			\bibnamefont {Yu}}, \ and\ \bibinfo {author} {\bibfnamefont {Qingming}\
			\bibnamefont {Zhang}},\ }\href@noop {} {\enquote {\bibinfo {title} {Pressure
				induced metallization and possible unconventional superconductivity in spin
				liquid \ce{NaYbSe2}},}\ } (\bibinfo {year} {2020}),\ \Eprint
	{http://arxiv.org/abs/2003.11479} {arXiv:2003.11479 [cond-mat.supr-con]}
	\BibitemShut {NoStop}%
	\bibitem [{\citenamefont {Jia}\ \emph {et~al.}(2020)\citenamefont {Jia},
		\citenamefont {Gong}, \citenamefont {Liu}, \citenamefont {Zhao},
		\citenamefont {Dong}, \citenamefont {Dai}, \citenamefont {Li}, \citenamefont
		{Lei}, \citenamefont {Yu}, \citenamefont {Zhang},\ and\ \citenamefont
		{Jin}}]{jia2020mott}%
	\BibitemOpen
	\bibfield  {author} {\bibinfo {author} {\bibfnamefont {Ya-Ting}\ \bibnamefont
			{Jia}}, \bibinfo {author} {\bibfnamefont {Chun-Sheng}\ \bibnamefont {Gong}},
		\bibinfo {author} {\bibfnamefont {Yi-Xuan}\ \bibnamefont {Liu}}, \bibinfo
		{author} {\bibfnamefont {Jian-Fa}\ \bibnamefont {Zhao}}, \bibinfo {author}
		{\bibfnamefont {Cheng}\ \bibnamefont {Dong}}, \bibinfo {author}
		{\bibfnamefont {Guang-Yang}\ \bibnamefont {Dai}}, \bibinfo {author}
		{\bibfnamefont {Xiao-Dong}\ \bibnamefont {Li}}, \bibinfo {author}
		{\bibfnamefont {He-Chang}\ \bibnamefont {Lei}}, \bibinfo {author}
		{\bibfnamefont {Run-Ze}\ \bibnamefont {Yu}}, \bibinfo {author} {\bibfnamefont
			{Guang-Ming}\ \bibnamefont {Zhang}}, \ and\ \bibinfo {author} {\bibfnamefont
			{Chang-Qing}\ \bibnamefont {Jin}},\ }\bibfield  {title} {\enquote {\bibinfo
			{title} {Mott transition and superconductivity in quantum spin liquid
				candidate \ce{NaYbSe2}},}\ }\href {\doibase 10.1088/0256-307x/37/9/097404}
	{\bibfield  {journal} {\bibinfo  {journal} {Chin. Phys. Lett.}\ }\textbf
		{\bibinfo {volume} {37}},\ \bibinfo {pages} {097404} (\bibinfo {year}
		{2020})}\BibitemShut {NoStop}%
	\bibitem [{SI()}]{SI}%
	\BibitemOpen
	\href@noop {} {\bibinfo  {journal} {Supplementary Information}\ }\BibitemShut
	{NoStop}%
	\bibitem [{\citenamefont {Li}\ \emph {et~al.}(2016{\natexlab{b}})\citenamefont
		{Li}, \citenamefont {Wang},\ and\ \citenamefont {Chen}}]{PhysRevB.94.035107}%
	\BibitemOpen
	\bibfield  {journal} {  }\bibfield  {author} {\bibinfo {author} {\bibfnamefont
			{Yao-Dong}\ \bibnamefont {Li}}, \bibinfo {author} {\bibfnamefont {Xiaoqun}\
			\bibnamefont {Wang}}, \ and\ \bibinfo {author} {\bibfnamefont {Gang}\
			\bibnamefont {Chen}},\ }\bibfield  {title} {\enquote {\bibinfo {title}
			{Anisotropic spin model of strong spin-orbit-coupled triangular
				antiferromagnets},}\ }\href {\doibase 10.1103/PhysRevB.94.035107} {\bibfield
		{journal} {\bibinfo  {journal} {Phys. Rev. B}\ }\textbf {\bibinfo {volume}
			{94}},\ \bibinfo {pages} {035107} (\bibinfo {year}
		{2016}{\natexlab{b}})}\BibitemShut {NoStop}%
	\bibitem [{\citenamefont {Schmidt}\ \emph {et~al.}(2021)\citenamefont
		{Schmidt}, \citenamefont {Sichelschmidt}, \citenamefont {Ranjith},
		\citenamefont {Doert},\ and\ \citenamefont {Baenitz}}]{PhysRevB.103.214445}%
	\BibitemOpen
	\bibfield  {author} {\bibinfo {author} {\bibfnamefont {B.}~\bibnamefont
			{Schmidt}}, \bibinfo {author} {\bibfnamefont {J.}~\bibnamefont
			{Sichelschmidt}}, \bibinfo {author} {\bibfnamefont {K.~M.}\ \bibnamefont
			{Ranjith}}, \bibinfo {author} {\bibfnamefont {Th.}\ \bibnamefont {Doert}}, \
		and\ \bibinfo {author} {\bibfnamefont {M.}~\bibnamefont {Baenitz}},\
	}\bibfield  {title} {\enquote {\bibinfo {title} {Yb delafossites: Unique
				exchange frustration of $4f$ spin-$\frac{1}{2}$ moments on a perfect
				triangular lattice},}\ }\href {\doibase 10.1103/PhysRevB.103.214445}
	{\bibfield  {journal} {\bibinfo  {journal} {Phys. Rev. B}\ }\textbf {\bibinfo
			{volume} {103}},\ \bibinfo {pages} {214445} (\bibinfo {year}
		{2021})}\BibitemShut {NoStop}%
	\bibitem [{\citenamefont {Li}\ \emph {et~al.}(2017)\citenamefont {Li},
		\citenamefont {Adroja}, \citenamefont {Bewley}, \citenamefont {Voneshen},
		\citenamefont {Tsirlin}, \citenamefont {Gegenwart},\ and\ \citenamefont
		{Zhang}}]{li2017crystalline}%
	\BibitemOpen
	\bibfield  {author} {\bibinfo {author} {\bibfnamefont {Yuesheng}\
			\bibnamefont {Li}}, \bibinfo {author} {\bibfnamefont {Devashibhai}\
			\bibnamefont {Adroja}}, \bibinfo {author} {\bibfnamefont {Robert~I}\
			\bibnamefont {Bewley}}, \bibinfo {author} {\bibfnamefont {David}\
			\bibnamefont {Voneshen}}, \bibinfo {author} {\bibfnamefont {Alexander~A}\
			\bibnamefont {Tsirlin}}, \bibinfo {author} {\bibfnamefont {Philipp}\
			\bibnamefont {Gegenwart}}, \ and\ \bibinfo {author} {\bibfnamefont
			{Qingming}\ \bibnamefont {Zhang}},\ }\bibfield  {title} {\enquote {\bibinfo
			{title} {Crystalline electric-field randomness in the triangular lattice
				spin-liquid \ce{YbMgGaO4}},}\ }\href {\doibase
		10.1103/PhysRevLett.118.107202} {\bibfield  {journal} {\bibinfo  {journal}
			{Phys. Rev. Lett.}\ }\textbf {\bibinfo {volume} {118}},\ \bibinfo {pages}
		{107202} (\bibinfo {year} {2017})}\BibitemShut {NoStop}%
	\bibitem [{\citenamefont {Sichelschmidt}\ \emph
		{et~al.}(2019{\natexlab{a}})\citenamefont {Sichelschmidt}, \citenamefont
		{Schlender}, \citenamefont {Schmidt}, \citenamefont {Baenitz},\ and\
		\citenamefont {Doert}}]{sichelschmidt2019electron}%
	\BibitemOpen
	\bibfield  {author} {\bibinfo {author} {\bibfnamefont {J{\"o}rg}\
			\bibnamefont {Sichelschmidt}}, \bibinfo {author} {\bibfnamefont {Philipp}\
			\bibnamefont {Schlender}}, \bibinfo {author} {\bibfnamefont {Burkhard}\
			\bibnamefont {Schmidt}}, \bibinfo {author} {\bibfnamefont {Michael}\
			\bibnamefont {Baenitz}}, \ and\ \bibinfo {author} {\bibfnamefont {Thomas}\
			\bibnamefont {Doert}},\ }\bibfield  {title} {\enquote {\bibinfo {title}
			{Electron spin resonance on the spin-$1/2$ triangular magnet \ce{NaYbS2}},}\
	}\href {\doibase 10.1088/1361-648x/ab071d} {\bibfield  {journal} {\bibinfo
			{journal} {J. Phys. Condens. Matter}\ }\textbf {\bibinfo {volume} {31}},\
		\bibinfo {pages} {205601} (\bibinfo {year} {2019}{\natexlab{a}})}\BibitemShut
	{NoStop}%
	\bibitem [{\citenamefont {Sichelschmidt}\ \emph {et~al.}(2020)\citenamefont
		{Sichelschmidt}, \citenamefont {Schmidt}, \citenamefont {Schlender},
		\citenamefont {Khim}, \citenamefont {Doert},\ and\ \citenamefont
		{Baenitz}}]{Sichelschmidt2020}%
	\BibitemOpen
	\bibfield  {author} {\bibinfo {author} {\bibfnamefont {Jörg}\ \bibnamefont
			{Sichelschmidt}}, \bibinfo {author} {\bibfnamefont {Burkhard}\ \bibnamefont
			{Schmidt}}, \bibinfo {author} {\bibfnamefont {Philipp}\ \bibnamefont
			{Schlender}}, \bibinfo {author} {\bibfnamefont {Seunghyun}\ \bibnamefont
			{Khim}}, \bibinfo {author} {\bibfnamefont {Thomas}\ \bibnamefont {Doert}}, \
		and\ \bibinfo {author} {\bibfnamefont {Michael}\ \bibnamefont {Baenitz}},\
	}\bibfield  {title} {\enquote {\bibinfo {title} {Effective spin-1/2 moments
				on a \ce{Yb^{3+}} triangular lattice: An {ESR} study},}\ }in\ \href {\doibase
		10.7566/jpscp.30.011096} {\emph {\bibinfo {booktitle} {Proceedings of the
				International Conference on Strongly Correlated Electron Systems
				({SCES}2019)}}}\ (\bibinfo  {publisher} {Journal of the Physical Society of
		Japan},\ \bibinfo {year} {2020})\BibitemShut {NoStop}%
	\bibitem [{\citenamefont {Sichelschmidt}\ \emph
		{et~al.}(2019{\natexlab{b}})\citenamefont {Sichelschmidt}, \citenamefont
		{Schlender}, \citenamefont {Schmidt}, \citenamefont {Baenitz},\ and\
		\citenamefont {Doert}}]{Sichelschmidt_2019}%
	\BibitemOpen
	\bibfield  {author} {\bibinfo {author} {\bibfnamefont {Jörg}\ \bibnamefont
			{Sichelschmidt}}, \bibinfo {author} {\bibfnamefont {Philipp}\ \bibnamefont
			{Schlender}}, \bibinfo {author} {\bibfnamefont {Burkhard}\ \bibnamefont
			{Schmidt}}, \bibinfo {author} {\bibfnamefont {Michael}\ \bibnamefont
			{Baenitz}}, \ and\ \bibinfo {author} {\bibfnamefont {Thomas}\ \bibnamefont
			{Doert}},\ }\bibfield  {title} {\enquote {\bibinfo {title} {Electron spin
				resonance on the spin-1/2 triangular magnet \ce{NaYbS2}},}\ }\href {\doibase
		10.1088/1361-648x/ab071d} {\bibfield  {journal} {\bibinfo  {journal} {Journal
				of Physics: Condensed Matter}\ }\textbf {\bibinfo {volume} {31}},\ \bibinfo
		{pages} {205601} (\bibinfo {year} {2019}{\natexlab{b}})}\BibitemShut
	{NoStop}%
\end{thebibliography}
%

\end{document}